\documentclass[a4paper,fleqn]{cas-sc}
\usepackage{diagbox}
\usepackage{algorithm}
\usepackage{algorithmicx}
\usepackage{algpseudocode}  
\makeatletter
\newenvironment{breakablealgorithm}
{
	\begin{center}
		\refstepcounter{algorithm}
		\hrule height.8pt depth0pt \kern2pt
		\renewcommand{\caption}[2][\relax]{
			{\raggedright\textbf{\ALG@name~\thealgorithm} ##2\par}%
			\ifx\relax##1\relax 
			\addcontentsline{loa}{algorithm}{\protect\numberline{\thealgorithm}##2}%
			\else 
			\addcontentsline{loa}{algorithm}{\protect\numberline{\thealgorithm}##1}%
			\fi
			\kern2pt\hrule\kern2pt
		}
	}{
		\kern2pt\hrule\relax
	\end{center}
}
\makeatother
\usepackage{subcaption}

\usepackage[authoryear,longnamesfirst]{natbib}

\def\tsc#1{\csdef{#1}{\textsc{\lowercase{#1}}\xspace}}
\tsc{WGM}
\tsc{QE}

\begin{document}
\let\WriteBookmarks\relax
\def\floatpagepagefraction{1}
\def\textpagefraction{.001}

\shorttitle{<short title of the paper for running head>}    

\shortauthors{<short author list for running head>}  

\title [mode = title]{Physics-informed Neural Network Combined with Characteristic-Based Split for Solving Navier-Stokes Equations}  

\tnotemark[1] 

\tnotetext[1]{This work was supported by the NSFC-Zhejiang Joint Fund for the Integration of Industrialization and Informatization under Grant U1809212 and the Fundamental Research Funds for the Central Universities under Grant xtr072022001.}

%

\author[1,2]{Shuang Hu}[]
\ead{11910077@zju.edu.cn}

\author[3,1]{Meiqin Liu}[]
\cormark[1]
\fnmark[1]
\ead{liumeiqin@zju.edu.cn}

\author[1,2]{Senlin Zhang}[]
\ead{slzhang@zju.edu.cn}

\author[1,2]{Shanling Dong}[]
\ead{shanlingdong28@zju.edu.cn}

\author[1,2]{Ronghao Zheng}[]
\ead{rzheng@zju.edu.cn}
\affiliation[1]{organization={College of Electrical Engineering},
	addressline={Zhejiang University}, 
	city={Hangzhou},
	postcode={310027}, 
	country={China}}
\affiliation[2]{organization={State Key Laboratory of Industrial Control Technology},
	addressline={Zhejiang University}, 
	city={Hangzhou},
	postcode={310027}, 
	country={China}}
\affiliation[3]{organization={Institute of Artificial Intelligence and Robotics},
	addressline={Xi’an Jiaotong University}, 
	city={Xi’an},
	postcode={710049}, 
	country={China}}
\cortext[1]{Corresponding author}

\fntext[1]{0000-0003-0693-6574}



\begin{abstract}
	In this paper, we propose a physics-informed neural network (PINN) combined with characteristic-based split (CBS) method for solving time-dependent Navier-Stokes equations (N-S equations). The proposed method separates the output parameters and corresponding losses, thereby disregarding weights between output parameters. Additionally, not all partial derivatives participate in gradient backpropagation, allowing for the reuse of remaining terms. As a result, this method is a faster version than traditional PINN. The labeled data, physical constraints, and network outputs are considered as priori information, and the residuals of the N-S equations are treated as posteriori information. This approach enables the handling of both data-driven and data-free problems. Notably, the method can solve the compressible N-S equations, specifically the Shallow-Water equations, and incompressible N-S equations. Given the known boundary conditions, the method requires only flow field information at a specific time to restore past and future flow field information. We demonstrate the potential of this method in marine engineering by solving the progress of a solitary wave onto a shelving beach and the dispersion of hot water in the flow. Furthermore, we prove the correctness and universality of the method by solving incompressible N-S equations with exact solutions. The source codes for the numerical examples in this work are available at \url{https://github.com/double110/PINN-cbs-.git}.
\end{abstract}


\begin{highlights}
\item This method disregards the weights between output parameters.
\item This method is a rapid version of Physics-informed Neural Network, as not all partial derivatives are involved in the gradient backpropagation, and the remaining terms are reused.
\item This method can solve Shallow-Water equations and incompressible N-S equations.
\item This method only requires the flow field information at a specific time to reconstruct the flow field information from both the past and future.
\end{highlights}


\maketitle

\section{Introduction}
Partial differential equations (PDEs) are mathematical equations that involve functions of multiple variables and their partial derivatives. They describe many physical phenomena, including the flow of fluids, the propagation of sound and light waves, and the diffusion of heat. PDEs are used extensively in physics, engineering, and other sciences to model complex systems and analyze their behavior.
The methods for solving PDEs based on neural networks are mostly data-driven. When there is sufficient spatiotemporal scattered point measurement data or the exact solution is known, using neural networks to solve computational fluid dynamics (CFD)-related problems such as parameter estimation, flow field reconstruction, and proxy model construction shows great potential.
\citet{ref5,ref6} proposed a feedforward neural network (PDE-Net) to invert unknown PDEs from data, in which the time derivative term is euler-discretized and the constrained convolution kernel approximates the differential operator. The method proposed in the work of \citet{ref7} can reconstruct the overall velocity field and pressure field with high resolution from sparse velocity information.
By adding a convection diffusion equation to N-S equations, the flow field can be restored from the concentration of the carrier in the fluid \citep{ref4}.

Although data-driven solutions have been widely used, they have certain limitations.  Methods that reduce label data and rely more on known equations are widely concerned.  Therefore, there is growing interest in methods that reduce the need for labeled data and rely more on known equations. One such method is the PINN proposed by \citet{ref1,ref2}. PINN constructs the residuals using the control equations of PDEs and the identity of the boundary conditions. The sum of the residuals is then used to construct the loss function. 
To further improve the performance of PINN, several modifications have been proposed. For example, \citet{ref8} combined PINN with orthogonal decomposition (POD) and discrete empirical interpolation method (DEIM) to provide a high-precision reduced order model of nonlinear dynamic systems and reduce the computational complexity of high-fidelity numerical simulations. Additionally, \citet{ref9} proposed the adaptive swish function to enhance the efficiency, robustness, and accuracy of PINN in approaching nonlinear functions and PDEs.  
\citet{ref19} implemented a finite-volume based numerical schemes inside the computational graph, which is based on CNN. \citet{ref20} proposed a distributed version of PINN, where the learning problem is decomposed into smaller regions of the computational domain and a physical compatibility condition is enforced in between neighboring domains. \citet{ref15} used the PINN framework composed of three DNNs to inverse the parameters of Richardson Richards equations, and achieved better results than one DNN. The multi-scale deep neural network (MscaleDNN) method was proposed to accelerate the convergence of high frequency \citep{ref13,ref14}, the core idea is to stretch the objective function at different scales in the radial direction. 

PINN has been proved to be able to solve multiple specification PDEs, and can handle multiple applications involving physics, chemistry and biology \citep{ref21,ref23,ref26,ref27,ref28,ref35}. PINN also has atypical applications. \citet{ref41} proposed a new framework for trajectory transfer optimization control, which obtains the first-order necessary conditions for the optimal control problem by applying the Pontryagin minimum principle. At this point, PINN transform solving partial differential problems into optimal control problems. So for logistics and transportation problems \citep{ref36,ref37,ref38,ref39,ref40}, the same solution approach can also be used. The model \citep{ref42} used assumes noisy measurements and a partially unknown first-order model to solve the problem of traffic density reconstruction using measurements from probe vehicles (PVs) with a low penetration rate. 
Using PINN framework to solve complex PDEs, such as N-S equations, is the focus of our work. \citet{ref29} investigated the possibility of using PINN to approximate the Euler equations that model high-speed aerodynamic flows. \citet{ref29} proposed Physics Informed Extreme Learning Machine (PIELM), which is a rapid version and  demonstrated to solve N-S equations in a lid-driven cavity at low Reynolds numbers.  \citet{ref19} proposed the DiscretizationNet, which employs a generative CNN-based encoder–decoder model with PDE variables as both input and output features. DiscretizationNet is demonstrated to solve the steady, incompressible N-S equations in 3-D for several cases such as, lid-driven cavity, flow past a cylinder and conjugate heat transfer.

The above mentioned method mainly optimizes PINN by modifying the network structure. Another type of method mainly focuses on processing the equations itself to adapt to th neural networks. By useing arbitrarily accurate implicit Runge–Kutta time stepping schemes with unlimited number of stages, \citet{ref30} proposed discrete time models. \citet{ref32} decomposed the order of differentiation to reduce complexity and avoid seeking high-order derivatives for a single neural network. Theory-guided hard constraint projection (HCP) \citep{ref33} used finite difference format templates to construct projection operators, and uses the objective function values obtained after projection as the target learning of the neural network. \citet{ref26} proposed the PhysGeoNet to solve the N-S equations using the finite difference discretizations of PDEs residuals in the neural network loss formulation.

The core idea of this paper is to assume that the current outputs of PINN are correct, and then investigate what the future and past flow fields should be. For this reason, we introduce CBS algorithm, which is widely used in the finite element method, into PINN. In the following examples, we only use the initial and boundary conditions to solve the equations without any labeled data.
There are two main types of N-S equations: compressible and incompressible. The primary difference between compressible and incompressible flows is the variation of density with pressure and temperature, which leads to different equations for their motion.
This method can solve  the special form of compressible N-S equations----Shallow-Water equations, and incompressible N-S equations. Compared with the traditional PINN, this method does not consider the weights of output variables, and the calculation cost is smaller. There are 23 buoys and 32 sonar buoys that can be publicly accessed within China. The accumulated oceanographic hydrological data has exceeded 50 million records since 2014. These data exhibit sparse spatial coverage but dense temporal resolution. Traditional spatial discretization methods cannot effectively utilize these data. Processing such data is what our method excels at. We show the progress of a solitary wave onto a shelving beach and the dispersion of the hot water in the flow to explain  the advantages of ocean flow field estimation. By solving incompressible equations with exact solutions, we prove this method's correctness and universality. 

\section{PINN based on CBS}
During the process of data fitting, neural networks often learn low frequencies before gradually learning high frequencies. This phenomenon is referred to as the Frequency Principle (F-Principle) \citep{ref11,ref12}. Gradient descent naturally aims to eliminate low-frequency errors, while high-dimensional errors cannot be eliminated. In fact, the F-Principle is also evident in the finite element method of N-S equations, such as the limitation of time step and grid density \citep{ref10}. The F-Principle reveals that high frequency disasters can occur in neural networks, and the training and generalization difficulties caused by these disasters cannot be easily alleviated through simple parameter adjustments.

\begin{figure}[h]
	\centering
	\begin{minipage}{.45\linewidth}
		\centering
		\includegraphics[scale=0.13]{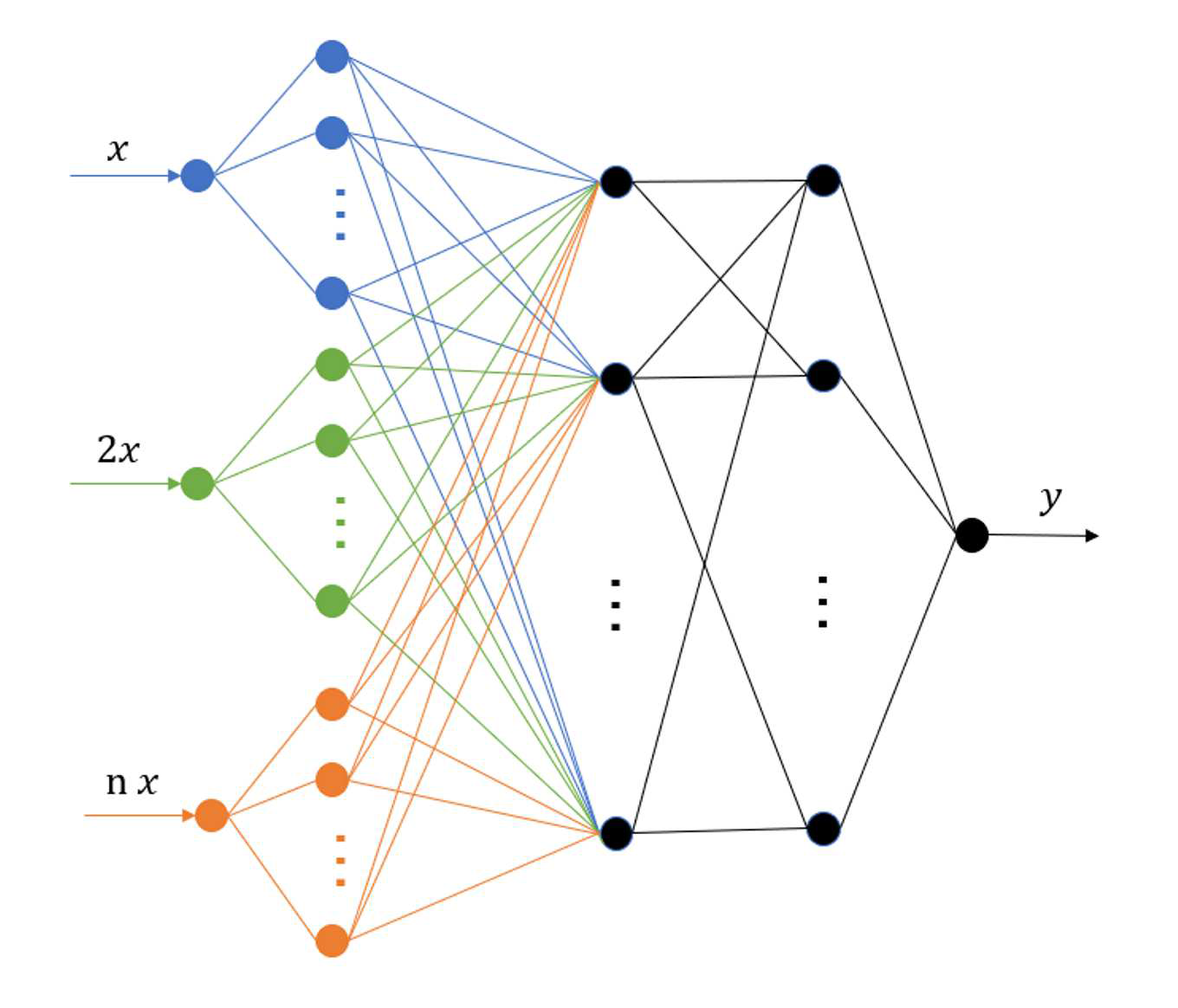}
		\centerline{$(a)$ MscaleDNN-1}
	\end{minipage}
	\hfill
	\begin{minipage}{.45\linewidth}
		\centering
		\includegraphics[scale=0.12]{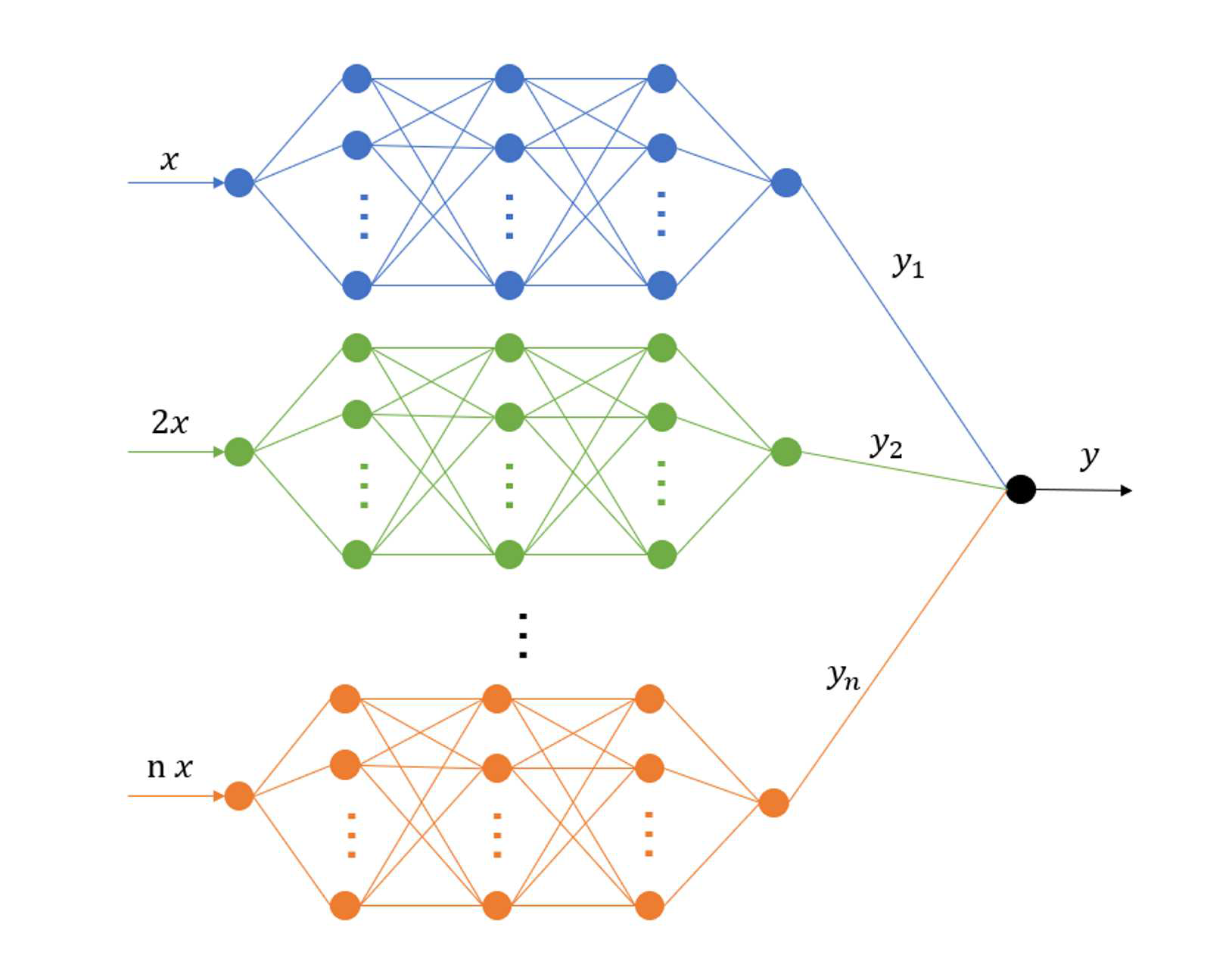}
		\centerline{$(b)$ MscaleDNN-2}
	\end{minipage}
	\caption{Two different network structures of MscaleDNN. The input set is a set of $ \textbf{x}=\left( x,2x,...,nx \right)$, and the output is $ y$.}
	\label{FIG:1}
\end{figure}

The MscaleDNN method can accelerate the convergence of high frequencies. The original MscaleDNN has two similar, but different, network structures, as shown in Figure \ref{FIG:1}. Both network structures are superior to fully connected networks. However, MscaleDNN-2, which appears to have a simpler structure, exhibits better performance than MscaleDNN-1. Before discussing why MscaleDNN-2 outperforms MscaleDNN-1, it is important to consider the following question: is it necessary to share parameters among output variables?

Suppose there exists a special solution of the N-S equations, where the velocity ${u}_{1}$ in the x direction varies with time, while the velocity ${u}_{2}$ in the y direction is constant. When ${u}_{1}$ and ${u}_{2}$ are fully connected, ${u}_{2}$ is not affected by time, so the weights coefficient of time in the full connection layer will be reduced. Meanwhile, ${u}_{1}$ is time-varying, and the weights reduction of time will cause the residuals related to ${u}_{1}$ to fail to converge, and vice versa. In fact, if we observe the partial derivative of the output ${u}_{1}$ and ${u}_{2}$ with respect to time under full connection, we will find that the two always remain at the same order of magnitude. This leads to the fact that the full connection between output parameters cannot converge in the case mentioned above, which also explains why MscaleDNN-2 is better than MscaleDNN-1. To some extent, MscaleDNN-2 achieves the separation of output parameters.

It is important to recognize that neural networks and the finite element method belong to the time-space approximation method. The finite element method, without specific algorithms, cannot solve complex problems, and similarly, the neural network also requires specific algorithms to achieve desirable results. After separating the output variables, we naturally consider an algorithm commonly used in the finite element method - CBS.

\subsection{Characteristic-based split algorithm}
The basic form of the N-S equations is as follows, and the shallow-water equations and incompressible-flow equations, which are mentioned later, are variant forms of it.

Conservation of mass:
\begin{equation}\label{eq2.1}
	\frac{\partial \rho }{\partial t}=\frac{1}{{{c}^{2}}}\frac{\partial p}{\partial t}=-\frac{\partial {{U}_{i}}}{\partial {{x}_{i}}}
\end{equation}

Conservation of momentum:
\begin{equation}\label{eq2.2}
	\frac{\partial {{U}_{i}}}{\partial t}=-\frac{\partial }{\partial {{x}_{j}}}\left( {{u}_{j}}{{U}_{i}} \right)+\frac{\partial {{\tau }_{ij}}}{\partial {{x}_{j}}}-\frac{\partial p}{\partial {{x}_{i}}}-{{Q}_{i}}
\end{equation}
where $i=direction$, $ {{U}_{i}}\text{=}\rho {{u}_{i}}$, $\rho$ is the density,  $p$ is the pressure, $c$ is the speed of sound, ${{u}_{i}}$ is the velocity component in the $i$ direction, and $T$ is the absolute temperature. $Q$ is body forces, and ${{Q}_{i}}$ is the component of $Q$ in the $i$ direction and the corrdinate axes are referred to as $x_{i}$. ${{\tau }_{ij}}$ is the deflection stress component, satisfying
\begin{equation}\label{eq2.3}
	{{\tau }_{ij}}\text{=}\mu \left( \frac{\partial {{U}_{i}}}{\partial {{x}_{j}}}\text{+}\frac{\partial {{U}_{j}}}{\partial {{x}_{i}}}-\frac{2}{3}{{\delta }_{ij}}\frac{\partial {{U}_{k}}}{\partial {{x}_{k}}} \right)
\end{equation}
where ${{\delta }_{ij}}$ is Kronecker delta. If $i=j$, ${{\delta }_{ij}}=1$; otherwise ${{\delta }_{ij}}=0$. $\mu $ is the viscosity coefficient.

For the convenience of calculation, we use the nondimensional form of each variable:
\begin{equation}
	\begin{aligned}
		& t=\frac{\bar{t}{{u}_{\infty }}}{L},{{x}_{i}}=\frac{{{{\bar{x}}}_{i}}}{L},\rho =\frac{\rho }{{{\rho }_{\infty }}},p=\frac{{\bar{p}}}{{{\rho }_{\infty }}u_{\infty }^{2}} \\ 
		& {{u}_{i}}=\frac{{{{\bar{u}}}_{i}}}{{{u}_{\infty }}},\operatorname{Re}=\frac{{{u}_{\infty }}L}{{{\nu }_{\infty }}},{{g}_{i}}=\frac{{{{\bar{g}}}_{i}}}{u_{\infty }^{2}},\nu =\frac{{\bar{\nu }}}{{{\nu }_{\infty }}} \\ 
	\end{aligned}
	\nonumber
\end{equation}

In the finite element calculation, we  discretize the equation in the time direction \cite{ref10}. Then Eq. \eqref{eq2.2} can be inferred as follows.
\begin{equation}\label{eq2.4}
	\begin{aligned}
		\frac{U_{i}^{n+1}-U_{i}^{n}}{\Delta t}= -\frac{\partial }{\partial {{x}_{j}}}{{\left( {{u}_{j}}{{U}_{i}} \right)}^{n}}+\frac{\partial \tau _{ij}^{n}}{\partial {{x}_{j}}}-\frac{\partial {{p}^{n+\theta }}}{\partial {{x}_{i}}}+Q_{i}^{n}
		+{{\left( \frac{\Delta t}{2}{{u}_{k}}\frac{\partial }{\partial {{x}_{k}}}\left( \frac{\partial }{\partial {{x}_{j}}}\left( {{u}_{j}}{{U}_{i}} \right)-\frac{\partial p}{\partial {{x}_{i}}}+{{Q}_{i}} \right) \right)}^{n}}
	\end{aligned}
\end{equation}
where ${{p}^{n+\theta }}$ represents the pressure value at time  $t={{t}^{n}}+\theta \Delta t$. $\Delta t$ represents the time step and $\theta \in \left( 0,1 \right)$. It further follows that
\begin{equation}\label{eq2.5}
	\begin{aligned}
		\frac{\partial {{p}^{n+\theta }}}{\partial {{x}_{i}}}=\theta \frac{\partial {{p}^{n+1}}}{\partial {{x}_{i}}}+\left( 1-\theta  \right)\frac{\partial {{p}^{n}}}{\partial {{x}_{i}}}\\
		\frac{\partial {{p}^{n-\theta }}}{\partial {{x}_{i}}}=\theta \frac{\partial {{p}^{n-1}}}{\partial {{x}_{i}}}+\left( 1-\theta  \right)\frac{\partial {{p}^{n}}}{\partial {{x}_{i}}}
	\end{aligned}
\end{equation}

At this stage, we utilize the CBS algorithm \cite{ref25} to substitute a suitable approximation, allowing for the calculation to be performed before obtaining ${{p}^{n+1}}$.

By introducing the auxiliary variables ${\Delta U}{i}^{*}$ and ${\Delta U}{i}^{**}$, we can split Eq. \eqref{eq2.4} into two parts
\begin{equation}\label{eq2.6}
	\frac{U_{i}^{n+1}-U_{i}^{n}}{\Delta t}=\frac{\Delta U_{i}^{*}}{\Delta t}+\frac{\Delta U_{i}^{**}}{\Delta t}
\end{equation}
\begin{equation}\label{eq2.7}
	\begin{aligned}
		\frac{\Delta U_{i}^{*}}{\Delta t}=-\frac{\partial }{\partial {{x}_{j}}}{{\left( {{u}_{j}}{{U}_{i}} \right)}^{n}}+\frac{\partial \tau _{ij}^{n}}{\partial {{x}_{j}}}+Q_{i}^{n} 
		+{{\left( \frac{\Delta t}{2}{{u}_{k}}\frac{\partial }{\partial {{x}_{k}}}\left( \frac{\partial }{\partial {{x}_{j}}}\left( {{u}_{j}}{{U}_{i}} \right)+{{Q}_{i}} \right) \right)}^{n}}
	\end{aligned}
\end{equation}
\begin{equation}\label{eq2.8}
	\frac{\Delta U_{i}^{**}}{\Delta t}=-\frac{\partial {{p}^{n+\theta_{2} }}}{\partial {{x}_{i}}}+\frac{\Delta t}{2}{{u}_{k}}\frac{{{\partial }^{2}}{{p}^{n}}}{\partial {{x}_{k}}\partial {{x}_{i}}}
\end{equation}

Eq. \eqref{eq2.7} is solved by an explicit time step applied to the discretized form. Eq. \eqref{eq2.8} is obtained once $\Delta p={{p}^{n+1}}-{{p}^{n}}$  is evaluated. From Eq. \eqref{eq2.1}, we have
\begin{equation}\label{eq2.9}
	\frac{\Delta \rho }{\Delta t}=\frac{1}{{{c}^{2}}}\frac{\Delta p}{\Delta t}=-\frac{\partial U_{i}^{n+\theta_{1}}}{\partial {{x}_{i}}}=-\left[ \frac{\partial U_{i}^{n}}{\partial {{x}_{i}}}+\theta_{1}\frac{\partial \Delta {{U}_{i}}}{\partial {{x}_{i}}} \right]
\end{equation}

Replacing  $\Delta {{U}_{i}}$ by $\Delta U_{i}^{*}$ , using Eq. \eqref{eq2.6} and Eq. \eqref{eq2.8} and rearranging and neglecting third- and higher-order terms, we obtain
\begin{equation}\label{eq2.10}
	\begin{aligned}
		\frac{\Delta \rho }{\Delta t}=\frac{1}{{{c}^{2}}}\frac{\Delta p}{\Delta t}=-\frac{\partial U_{i}^{n}}{\partial {{x}_{i}}}-\theta_{1}\frac{\partial \Delta U_{i}^{*}}{\partial {{x}_{i}}}
		+\Delta t\theta_{1}\frac{{{\partial }^{2}}{{p}^{n+{{\theta }_{2}}}}}{\partial {{x}_{i}}\partial {{x}_{i}}}
	\end{aligned}
\end{equation}

In the equations above, $0.5\le {{\theta }_{1}}\le 1.0$ and ${\theta }_{2}=0$ for explicit scheme, while $0.5\le {{\theta }_{1}}\le 1.0$ and $0.5\le {{\theta }_{2}}\le 1.0$  for the semi-implicit scheme. In the following calculation process, we choosed the  semi-implicit scheme with ${\theta }_{1}={\theta }_{2}=0.5$. The governing equations can be solved after spatial discretization in the following order:

\begin{itemize}
	\item Step 1: use Eq. \eqref{eq2.7} to obtain $\Delta U_{i}^{*}$. 
	\item Step 2: use Eq. \eqref{eq2.10} to obtain $\Delta p$. 
	\item Step 3: use Eq. \eqref{eq2.8} to obtain $\Delta U_{i}^{**}$ thus establishing the values of $U_{i}$ and $p$.
\end{itemize}

After completing the calculation to establish $\Delta U_{i}$ and $\Delta p$, the transport equation is handled independently, as discussed in Section \ref{Sec3.3}.

The CBS algorithm was initially used in the finite element method, and it retains second-order accuracy in the calculation process. While it is not impossible to achieve higher-order accuracy using neural networks, the amount of calculation required increases exponentially. Second-order accuracy has been demonstrated to be accurate enough in the finite difference method \cite{ref18}; therefore, no higher-order CBS algorithm is derived in this study.

\subsection{Physics-informed neural network combined with characteristic-based split method}
A PINN is essentially a DNN that can be used to approximate the solution determined by the data and PDEs. A residual neural network can be expressed as
\begin{equation}\label{eq2.11}
	\left( \textbf{u},p \right)={{F}_{NN}}\left( \textbf{x},t;\Theta  \right)
\end{equation}
where ${F}_{NN}$ represents the neural network, whose inputs are space coordinates $ \textbf{x}=\left( {{x}_{1}},{{x}_{2}},{{x}_{3}} \right)$ and time $t$. The parameter $\Theta$ represents the trainable variables. The outputs of the neural network are velocity vector $ \textbf{u}=\left( {{u}_{1}},{{u}_{2}},{{u}_{3}} \right)$ and pressure $p$. We set up separate networks for each output and use independent optimizer and learning rate for each network. In the process of selecting sampling points, we think the sampling points should meet certain requirements. In the finite element method, there must be at least ten points in a wavelength to correctly describe the waveform without divergence. Random sampling does not necessarily ensure that convergence requirements are satisfied everywhere in the space. If the random sampling result causes divergence, it will be better to fix the sampling points. The examples selected in the following sections are all in two-dimensional rectangular space, so the sampling points are selected as the structured grid. In fact, the unstructured mesh in the finite element method can also be selected as the sampling points. The time interval $\Delta t$ is fixed. The calculation steps are as follows:
\begin{figure*}[htbp]
	\centering
	\includegraphics[width=0.9\columnwidth]{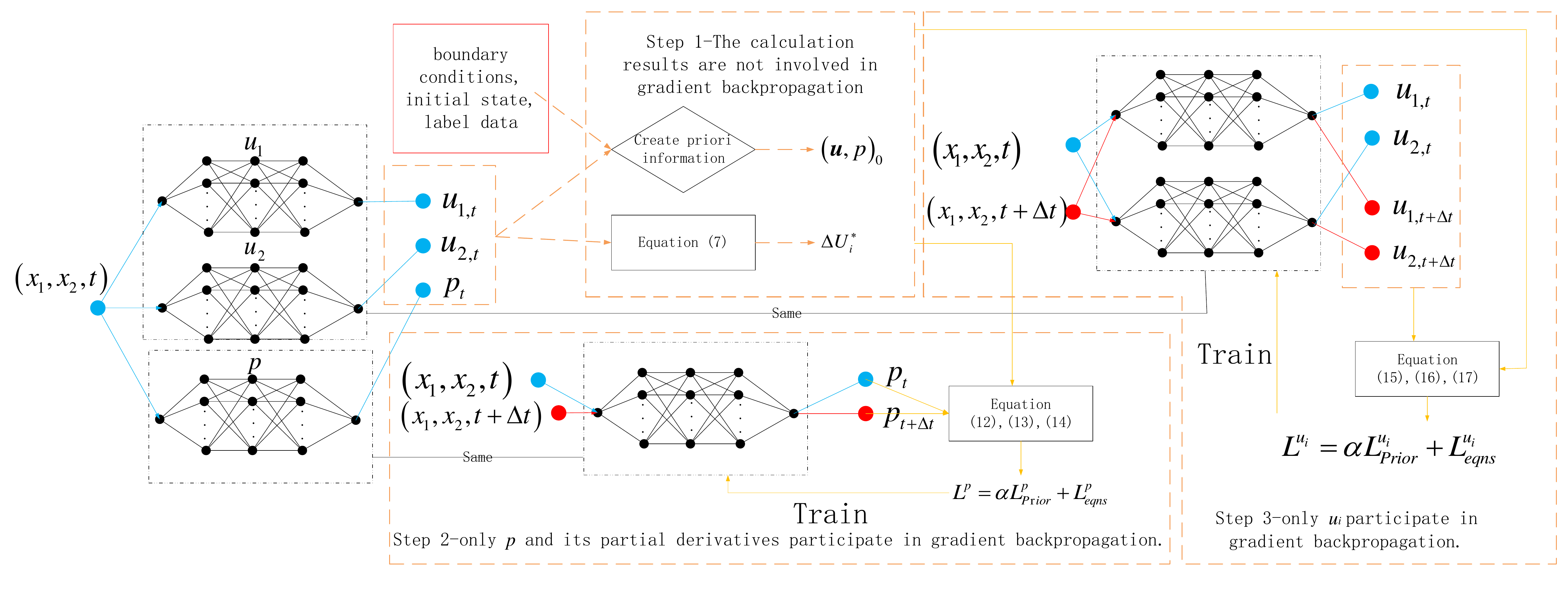}
	\caption{The solution process of PINN combined with CBS for solving N-S equations. The inputs are space coordinates $ \textbf{x}=\left( {{x}_{1}},{{x}_{2}},{{x}_{3}} \right)$ and time $t$, and the outputs are velocity vector $ \textbf{u}=\left( {{u}_{1}},{{u}_{2}},{{u}_{3}} \right)$ and pressure $p$. The physical laws are represented by N–S equations and automatic differentiation operators are used to obtain partial derivatives.}
	\label{FIG:2}
\end{figure*}

Step 1,

Use input N sampling points $\left( \textbf{x},t \right)^{j}$ to obtain the initial output $\left( \textbf{u},p \right)^{j}_{b}$. $\Delta U_{i}^{*}$ can be calculated by Eq. \eqref{eq2.7}. We consider $\left( \textbf{u},t \right)^{j}_{b}$ as prior information,  similar to boundary conditions, initial state and label data. Thus, we combine $\left( \textbf{u},t \right)^{j}_{b}$, boundary conditions, initial state and label data together to construct prior information $\left( \textbf{u},p \right)^{j}_{0}$. The construction method involves replacing the original output point with the boundary conditions, initial state, and label data.

Step 2,

Use input N sampling points $\left( \textbf{x},t+\Delta t \right)^{j}$ to obtain $p_{a}^{j}$. $\Delta p_{a}^{j}$ can be calculated by Eq. \eqref{eq2.10}. To make the network satisfy the governing equations, the loss function in the $p$'s network is defined as follows:

\begin{equation}\label{eq2.12}
	{{L}^{p}}=\alpha^{p} L_{Prior}^{p}+L_{eqns}^{p}
\end{equation}
where $\alpha^{p}$ is a weighting coefficient. $L_{Prior}^{p}$ and $L_{eqns}^{p}$ are computed as

\begin{equation}\label{eq2.13}
	L_{Prior}^{p}=\sum\limits_{j=1}^{N}{{{\left( {{p}_{0}^{j}}-{{p}_{b}^{j}} \right)}^{2}}/\Delta{{t}^{2}}}
\end{equation}
\begin{equation}\label{eq2.14}
	L_{eqns}^{p}=\sum\limits_{j=1}^{N}{{{\left( {{p}_{a}^{j}}-{{p}_{0}^{j}}-\Delta {{p}_{a}^{j}} \right)}^{2}}/\Delta{{t}^{2}}}
\end{equation}
where $L_{Prior}^{p}$ represents the loss between the prior information and the predicted data and $L_{eqns}^{p}$ denotes the total residual of the N-S equations. Note that we use $p_{0}$ instead of $p_{b}$ as the target result. This is also a common practice in the finite element method. Boundary conditions or known conditions replace the original data in the calculation process rather than after the calculation, which can obtain higher accuracy. Note that only $p_{a}$ and its partial derivatives participate in gradient backpropagation.

Repeat Step 2 for $K^{p}$ times and recalculate  $\Delta p_{a}^{j}$.

Step 3,

Use input N sampling points $\left( \textbf{x},t+\Delta t \right)^{j}$ to obtain $u_{i,a}^{j}$. $\Delta U_{i,a}^{j}$ can be calculated by Eq. \eqref{eq2.8}. $\rho_{0}$ and $\rho_{a}$ can be calculated by Eq. \eqref{eq2.10}. The loss function in the $u_{i}$'s network is defined as follows:
\begin{equation}\label{eq2.15}
	{{L}^{u_{i}}}=\alpha^{u_{i}} L_{Prior}^{u_{i}}+L_{eqns}^{u_{i}}
\end{equation}
where $\alpha^{u_{i}}$ is a weighting coefficient, and $L_{Prior}^{p}$ and $L_{eqns}^{p}$ are computed as
\begin{equation}\label{eq2.16}
	L_{Prior}^{u_{i}}=\sum\limits_{j=1}^{N}{{{\left( {{u_{i,0}}^{j}}-{{u_{i,b}}^{j}} \right)}^{2}}/\Delta{{t}^{2}}}
\end{equation}
\begin{equation}\label{eq2.17}
	L_{eqns}^{{{u}_{i}}}=\sum\limits_{j=1}^{N}{{{\left( {{u}_{i,a}}^{j}-\left( {{\rho }_{0}}{{u}_{i,0}}^{j}+\Delta {{U}_{i,a}}^{j} \right)/{{\rho }_{a}} \right)}^{2}}/\Delta {{t}^{2}}}
\end{equation}
where $L_{Prior}^{u_{i}}$ represents the loss between the prior information and the predicted data, and $L_{eqns}^{u_{i}}$ denotes the total residual of the N-S equations. Note that only $u_{i,a}$ participate in gradient backpropagation.

Repeat Step 3 for $K^{u_{i}}$ times and retain  $\Delta U_{i,a}^{j}$.

Repeat Step 1-3 until $L$ convergences to required accuracy, shown in Fig. \ref{FIG:2}. It can be seen that each input point in each step has two reference data: one is the prior information $I_{0}$, and the other is the posterior information $I_{1}$ obtained from Eq. \eqref{eq2.1}-\eqref{eq2.2}. Therefore, theoretically, the final output result of each step will converge to $I_{final}=(I_{1}+\alpha I_{0})/(1+\alpha)$. And the final loss $L_{final}$ is related to the initial loss $L_{init}$ as $L_{final} >= \frac{\alpha}{(1+\alpha)} L_{init} $. Hence, when loss decreases by  $\frac{1}{2(1+\alpha)} L_{init}$, there is no need to increase $K$. When $\alpha$ decreases, the convergence speed will accelerate, but it is more likely to diverge.If the output continues to diverge, we should increase the corresponding $\alpha$. Initially, $\alpha$ should be set to a larger value to prevent divergence, and then adjusted to a smaller value after stabilizing to speed up convergence. As $\alpha$ may change during the training process, the values of $\alpha$ mentioned below will represent the last set values.

It should be noted that the theoretical convergence result of this method does not depend on the network structure. When the network can fit the results with the required accuracy, changing the network structure will only affect the convergence rate. Therefore, all the networks mentioned below have the same network structure: the number of hidden layers $N_{layer}=6$, the number of neurons $N_{cell}=128$ in each hidden layer, and the activation function is the Swish function $\sigma (x)=\frac{x}{1+e^{-x}}$. To compute the residuals of the N-S equations, the partial differential operators are calculated using automatic differentiation (AD), which can be directly formulated in the deep learning framework, such as using “torch.autograd.grad” in Torch. In AD, the derivatives in the governing equations are approximated by the derivatives of the output with respect to the input of the PINNs. The use of “with torch.no\_ Grad()” indicates that the current calculation does not require backpropagation.

The following is divided into two parts: 1. Use the shallow-water equations to solve the solitary wave problem, and it is pointed out that the separation of output parameters is more suitable for practical engineering applications; and 2. Solve the incompressible N-S equations with exact solutions and prove the method's correctness and universality.

\subsection{Autograd mechanics and backpropagation}

Our work is conducted within the framework of PyTorch, and we will briefly introduce its autograd mechanics and backpropagation mechanism, which complement each other. PyTorch is an open-source machine learning library that is primarily used for developing deep learning models. PyTorch provides a dynamic computational graph, which allows for easy debugging and flexibility in model design. The following analysis is based on the official PyTorch documentation, available at \url{https://pytorch.org/docs/2.0/notes/autograd.html#autograd-mechanics}. 

Each variable in PyTorch has a $creator$ attribute that points to the function that takes it as output. This is the entry point of a directed acyclic graph (DAG) composed of Function objects as nodes. The reference between them represents the edge of the graph. Every time an operation is executed, a new function representing it is instantiated, its $forward()$ method is called, and the $creator$ of the variable it outputs is set to this function. By tracking the path from any variable to the leaf node, the sequence of operations used to generate the data can be reconstructed, and the gradient can be automatically calculated.

\begin{figure*}[htbp]
	\centering
	\includegraphics[width=0.9\columnwidth]{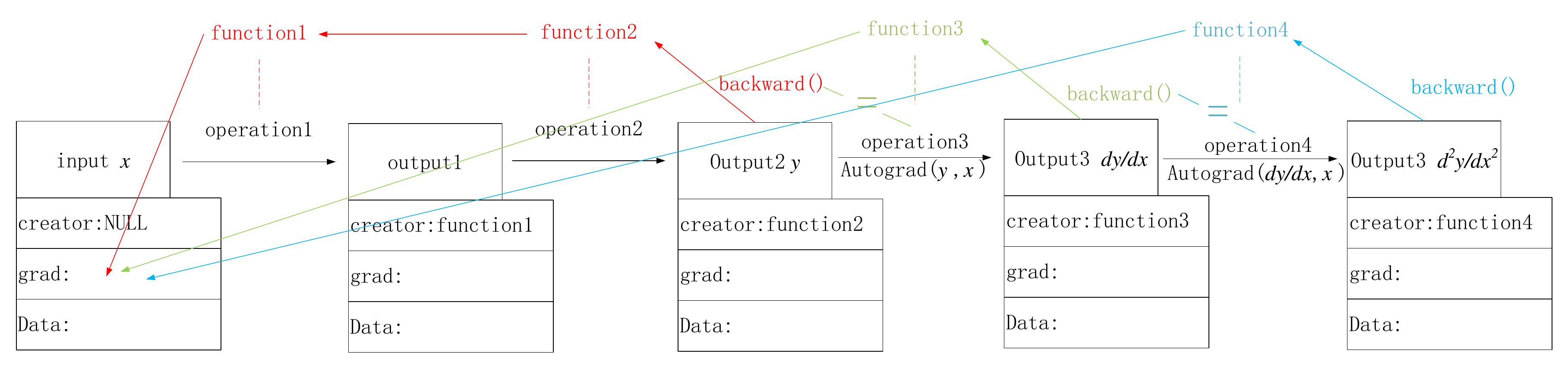}
	\caption{The brief process of autograd mechanics.}
	\label{FIG:11}
\end{figure*}

As depicted in Fig. \ref{FIG:11}, let us assume that we have an input variable $x$ (with a data type of Variable) that is provided by the user, and thus its $creator$ is NULL. After the first data operation, operation1 (such as addition, subtraction, multiplication, and division), the output1 variable (with a data type of Variable) is generated. In this process, a new instance of function1 (with a data type of Function) is automatically created, and the $creator$ of output1 is set to this function1. Subsequently, output1 undergoes another data operation to generate output2, which also generates another instance of function2. The $creator$ of output2 $y$ is function2.

\begin{table}[!h]
	\centering
	\caption{The memory occupation increment as the order of differentiation increases under autograd mechanics. The neural network architecture is 9 layers and 20 neurons per layer. The training data is $N=20000$.}\label{tbl1}
	\begin{tabular}{ c|cccc }
		\hline
		Derivative order &  $dy/dx$ & $d^{2}y/dx^{2}$ & $d^{3}y/dx^{3}$ & $d^{4}y/dx^{4}$  \\
		\hline
		memory(Mb) & 48 & 120 & 280 & 640 \\
		\hline
	\end{tabular}
\end{table}

During the forward propagation process, function1 and function2 record the operation history of the input $x$. When $y$ runs $backward()$, it causes function2 and function1 to automatically calculate the derivative value of $x$ in reverse and store it in the $grad$ attribute. At the same time, the graph is destroyed, and memory is reclaimed. When using $Autograd(y, x)$ to calculate $dy/dx$, it is equivalent to using $backward()$ without destroying the graph and generating a new function3. Due to the involvement of function2, function1 and their derivatives in function3, function3 occupies much more memory than the sum of function2 and function1. Moreover, its $backward()$ consumes more computing power. Similarly, when using $Autograd(dy/dx, x)$ to calculate $d^{2}y/dx^{2}$, function4 is generated and occupies more memory than function3. For instance, if a single-input single-output network has a network structure of 9 layers and 20 neurons per layer, and the training data is $N=20000$, the memory occupation increment as the order of differentiation increases is shown in Table \ref{tbl1}.

As we can see, the memory occupation and the computing power required for backpropagation both increase exponentially as the order of differentiation increases. Therefore, by blocking the backpropagation of higher-order partial derivatives, we can significantly reduce the consumption of memory and computing power by destroying the corresponding function and preventing backward calculation.

\section{Shallow Water Problems}
\label{Sec3}
The shallow-water equations are a set of partial differential equations that describe the behavior of fluids with a small depth compared to their horizontal extent. They are widely used to model the motion of fluids in oceans, lakes, and other bodies of water, as well as the flow of air in the atmosphere.
The isothermal compressible folw equations can be transformed into the depth integrated shallow-water equations with the variables being changed as follows:
\[ 
\begin{aligned}
	& \rho \left( density \right)\to h\left( depth \right) \\ 
	& {{u}_{i}}\left( velocity \right)\to {{{\bar{u}}}_{i}}\left( mean-velocity \right) \\ 
	& p\left( pressure \right)\to \frac{1}{2}g\left( {{h}^{2}}-{{H}^{2}} \right)
\end{aligned}  
\]

It can be written in a convenient form for the general CBS formulation as
\begin{equation}\label{eq3.1}
	\frac{\partial h}{\partial t}+\frac{\partial {{U}_{i}}}{\partial {{x}_{i}}}=0
\end{equation}
\begin{equation}\label{eq3.2}
	\frac{\partial {{U}_{i}}}{\partial t}+\frac{\partial \left( {{{\bar{u}}}_{j}}{{U}_{i}} \right)}{\partial {{x}_{j}}}+\frac{\partial }{\partial {{x}_{i}}}\left( \frac{1}{2}g\left( {{h}^{2}}-{{H}^{2}} \right) \right)+{{Q}_{i}}=0
\end{equation}
where $U_{i}=h{{\bar{u}}}_{i}$, and $H$ is the mean-depth. The rest of the variables are the same as those described before. Set $Q_{i}=0$. The three essential steps of the CBS scheme can be written in its semi-discrete form as

Step 1,
\begin{equation}\label{eq3.3}
	\begin{aligned}
		\Delta U_{i,a}^{*}= 
		\Delta t{{\left[ -\frac{\partial \left( {{{\bar{u}}}_{j}}{{U}_{i}} \right)}{\partial {{x}_{j}}}
				+\frac{\Delta t}{2}{{u}_{k}}\frac{\partial }{\partial {{x}_{i}}}\left( \frac{\partial \left( {{{\bar{u}}}_{j}}{{U}_{i}} \right)}{\partial {{x}_{j}}} \right) \right]}^{n}}\\
	\end{aligned}
\end{equation}

Step 2,
\begin{equation}\label{eq3.4}
	\Delta {{h}_{a}}=-\Delta t{{\left[ \frac{\partial \left( U_{i}^{n} \right)}{\partial {{x}_{j}}}+{{\theta }_{1}}\frac{\partial \Delta U_{i}^{*}}{\partial {{x}_{i}}}-\Delta t{{\theta }_{1}}\frac{{{\partial }^{2}}{{p}^{n+{{\theta }_{2}}}}}{\partial {{x}_{i}}\partial {{x}_{i}}} \right]}}
\end{equation}
\begin{equation}\label{eq3.5}
	L_{Prior}^{h}=\sum\limits_{j=1}^{N}{{{\left( {{h}_{0}}-{{h}_{b}} \right)}^{2}}/{{\Delta t}^{2}}}
\end{equation}	
\begin{equation}\label{eq3.6}
	L_{eqns}^{h}=\sum\limits_{j=1}^{N}{{{\left( {{h}_{a}}-{{h}_{\text{0}}}-\Delta {{h}_{a}} \right)}^{2}}/{{\Delta t}^{2}}}
\end{equation}	

Step 3,
\begin{equation}\label{eq3.7}
	\Delta {{U}_{i,a}}=\Delta U_{i,a}^{*}-\Delta t\frac{\partial p_{i}^{n+{{\theta }_{2}}}}{\partial {{x}_{i}}}+\frac{\Delta {{t}^{2}}}{2}{{\bar{u}}_{k}}\frac{{{\partial }^{2}}{{p}^{n}}}{\partial {{x}_{k}}\partial {{x}_{i}}}
\end{equation}	
\begin{equation}\label{eq3.8}
	L_{Prior}^{{{u}_{i}}}=\sum\limits_{j=1}^{N}{{{\left( {{u}_{i,0}}^{j}-{{u}_{i,b}}^{j} \right)}^{2}}/\Delta {{t}^{2}}}
\end{equation}	
\begin{equation}\label{eq3.9}
	L_{eqns}^{{{u}_{i}}}=\sum\limits_{j=1}^{N}{{{\left( {{u}_{i,a}}^{j}-\left( {{h}_{0}}{{u}_{i,0}}^{j}+\Delta {{U}_{i,a}}^{j} \right)/{{h}_{a}} \right)}^{2}}/\Delta {{t}^{2}}}\
\end{equation}	
with $p=g(h^{2}-H^{2})$. 

After Step 1, the calculation graph generated in this step will be destroyed and only the value of $\Delta U_{i}^{*}$ will be retained. This step does not involve backpropagation. After Step 2, only the calculation graphs that generate $\frac{{{\partial }^{2}}{p}}{\partial {{x}_{i}}\partial {{x}_{i}}}$ and $h$ are retained. After Step 3, only the calculation graph of $u_{i}$ is retained. Only parameters related to the retained computational graph can participate in backpropagation.

Solitary waves, also known as solitons, are localized waves that maintain their shape and speed as they propagate through a medium. Unlike most waves, which disperse and lose energy as they travel, solitary waves can travel long distances without changing their shape or size. Solitary waves have many practical applications, including in optical communications, where they can be used to transmit information without distortion or attenuation, and in oceanography, where they can help to model and predict the behavior of ocean waves.

The propagation process of a solitary wave in a flat-bottomed flume has an accurate solution, which can be used to verify the correctness of the method. In this case, the wave height and velocity of solitary waves do not change during the propagation process in the flat-bottomed flume.

\begin{figure}[h]
	\centering
	\includegraphics[width=0.5\columnwidth]{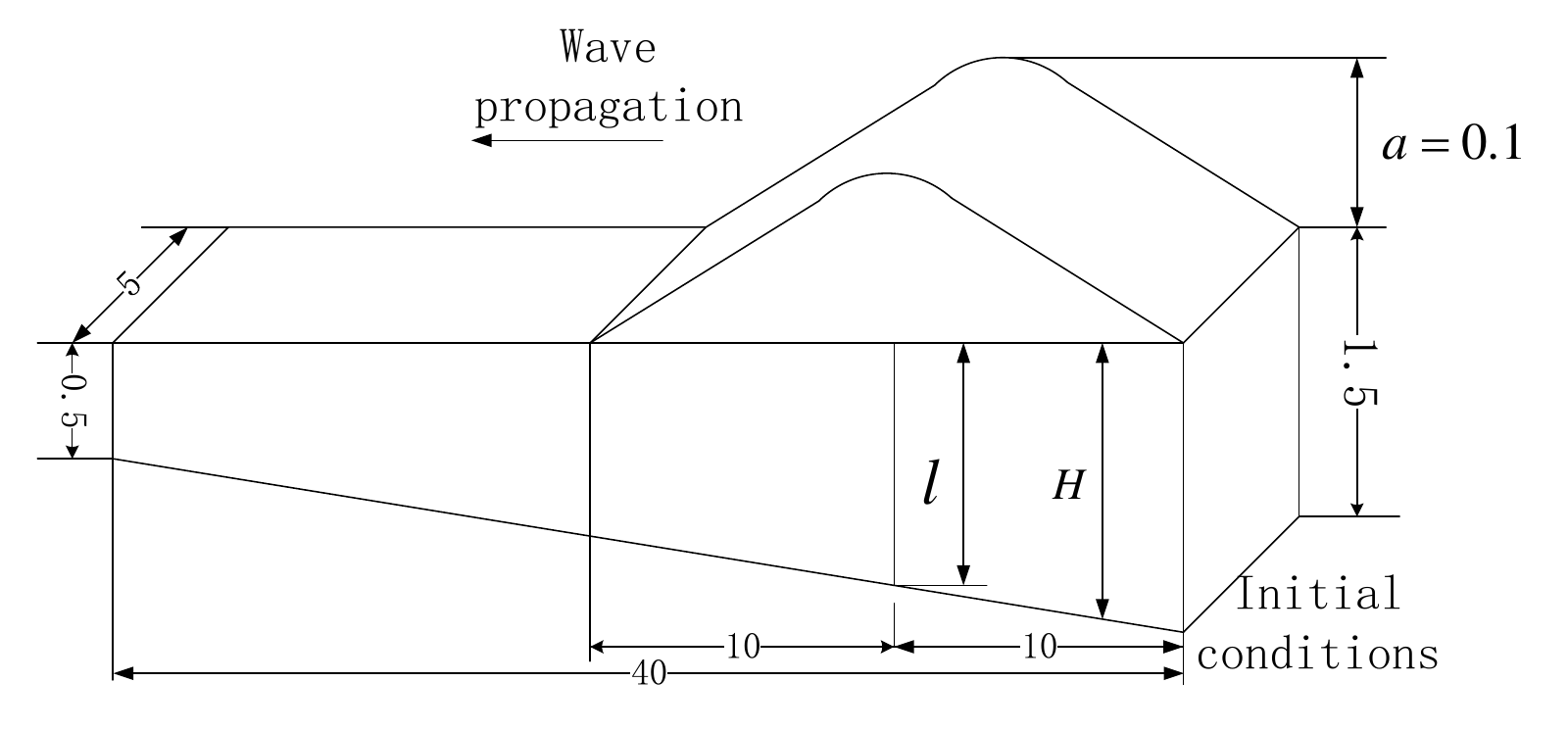}
	\caption{Solitary wave onto a shelving beach. }
	\label{FIG:3}
\end{figure}

To make the problem more complex, we can compute the propagation process of solitary waves onto a shelving beach, as shown in Fig. \ref{FIG:3}. The propagation process of a solitary wave \cite{ref16} can be approximately expressed as:
\begin{equation}\label{eq3.10}
	\begin{aligned}
		&\eta(x_{1},x_{2},t) =a{{\operatorname{sech}}^{2}}\left( \sqrt{3a/4{{h}^{3}}}\times \left( x_{1}-30-Ct \right) \right)\\
		&u_{1}(x_{1},x_{2},t)=-\left( l+1/2a \right)\eta /\left( (x_{1}-Ct)/30+\eta  \right)
	\end{aligned}
\end{equation}	
with $u_{2}(x,y,t)=0$, $a=0.1$, $\eta =h-H$ and $C=\sqrt{gh\left( 1+H/h \right)}$. Set $g=1.0$, so $C\approx 1.0$. It should be pointed out that the above equation is its approximate expression. But regardless of whether the equation is correct or not, this method can always derive the solution that meets the priori information and shallow-water equations.

\subsection{Solve when the initial state is known}
The computation is carried out in the domain of $0\le x_{1}\le 40$, $0\le x_{2}\le 5 $ and $0\le t\le 10 $. Sampling interval is $\Delta x_{1}=0.25$, $\Delta x_{2}=0.25$
and $\Delta t=0.125$. So the total number of sampling points is $N=161*21*81=273861$.
All three networks use the Adam optimizer and $\alpha^{p}=\alpha^{u_{1}}=\alpha^{u_{2}}=1$. The initial state equations and boundary conditions are
\begin{equation}\label{eq3.12}
	\begin{aligned}
		&\eta(x_{1},x_{2},0) =a{{\operatorname{sech}}^{2}}\left( \sqrt{3a/4{{h}^{3}}}\times \left( x_{1}-30 \right) \right)\\
		&u_{1}(x_{1},x_{2},0)=-\left( l+1/2a \right)\eta /\left( x_{1}/30+\eta  \right)
	\end{aligned}
\end{equation}	
\begin{equation}\label{eq3.14}
	\left\{ \begin{aligned}
		& u_{1}\left( 0,{{x}_{2}},t \right)=0 , u_{1}\left( 40,{{x}_{2}},t \right)=0 \\ 
		& u_{2}\left( {{x}_{1}},0,t \right)=0 , u_{2}\left( {{x}_{1}},5,t \right)=0 \\ 
	\end{aligned} \right.
\end{equation}	

The pseudo code of the calculation process is as follows:
\begin{breakablealgorithm}
	\caption{shallow-water problem }
	\begin{algorithmic}[1]
		\State Set $\Delta {{t}}=0.125$
		\State get $h_{ini},\bar{u}_{i,ini}$ by Eq. \eqref{eq3.12}
		\For{step in range(1000)} 
		
		\State //Step 1 
		\State input N points $\left( \textbf{x},t \right)$ to obtain $\bar{u}_{i,b},h_{b}$
		\State with torch.no\_grad(): \Comment{Create priori information}
		\State ~~~~$h_{0},\bar{u}_{i,0}=clone(\bar{u}_{i,b},h_{b})$
		\State ~~~~$h_{0}(:,t=0),\bar{u}_{i,0}(:,t=0)=h_{ini},\bar{u}_{i,ini}$
		\State ~~~~$\bar{u}_{1,0}(x_{2}=0,:)=\bar{u}_{1,0}(x_{2}=40,:)=0$
		\State ~~~~$\bar{u}_{2,0}(x_{1}=0,:)=\bar{u}_{2,0}(x_{1}=5,:)=0$
		\State get $\Delta U_{i,a}^{*}$ by Eq. \eqref{eq3.3}.     
		
		\State //Step 2 
		\State with torch.no\_grad(): \Comment{Prepare for $\Delta {{h}_{a}}$}
		\State ~~~~get $\frac{\partial \left( U_{i}^{n} \right)}{\partial {{x}_{j}}},\frac{\partial \Delta U_{i}^{*}}{\partial {{x}_{i}}}$ by AD
		\For{$k$ in range($K_{h}$)}
		\State input N points $\left( \textbf{x},t \right)$ and $\left( \textbf{x},t+\Delta t \right)$ to obtain $h_{b}$ and $h_{a}$, Respectively 
		\State get $\frac{{{\partial }^{2}}{{p}^{n+{{\theta }_{2}}}}}{\partial {{x}_{i}}\partial {{x}_{i}}}$ \Comment{Gradient is not mandatory}
		\State get $\Delta {{h}_{a}}$ by Eq. \eqref{eq3.4}
		\State get $L^{h}$ by Eqs. \eqref{eq3.5},\eqref{eq3.6}
		\State $L^{h}.backward()$ and use Adam Optimizer 
		\EndFor
		
		\State //Step 3
		\State with torch.no\_grad(): 
		\State ~~~~input N points $\left( \textbf{x},t+\Delta t \right)$ to obtain $h_{a}$
		\State ~~~~get $\frac{\partial p_{i,a}}{\partial {{x}_{i}}}$
		\State ~~~~get $\Delta {{U}_{i,a}}$ by Eq. \eqref{eq3.7}
		\For{$k$ in range($K_{u_{i}}$)}
		\State input N points $\left( \textbf{x},t \right)$ and $\left( \textbf{x},t+\Delta t \right)$ to obtain $\bar{u}_{i,b}$ and $\bar{u}_{i,a}$, Respectively
		\State get $L^{u_{i}}$ by Eqs. \eqref{eq3.8},\eqref{eq3.9}
		\State $L^{u_{i}}.backward()$ and use Adam Optimizer 
		\EndFor
		\EndFor
	\end{algorithmic}
\end{breakablealgorithm}

\begin{figure*}[htb]
	\centering
	\includegraphics[width=0.70\columnwidth]{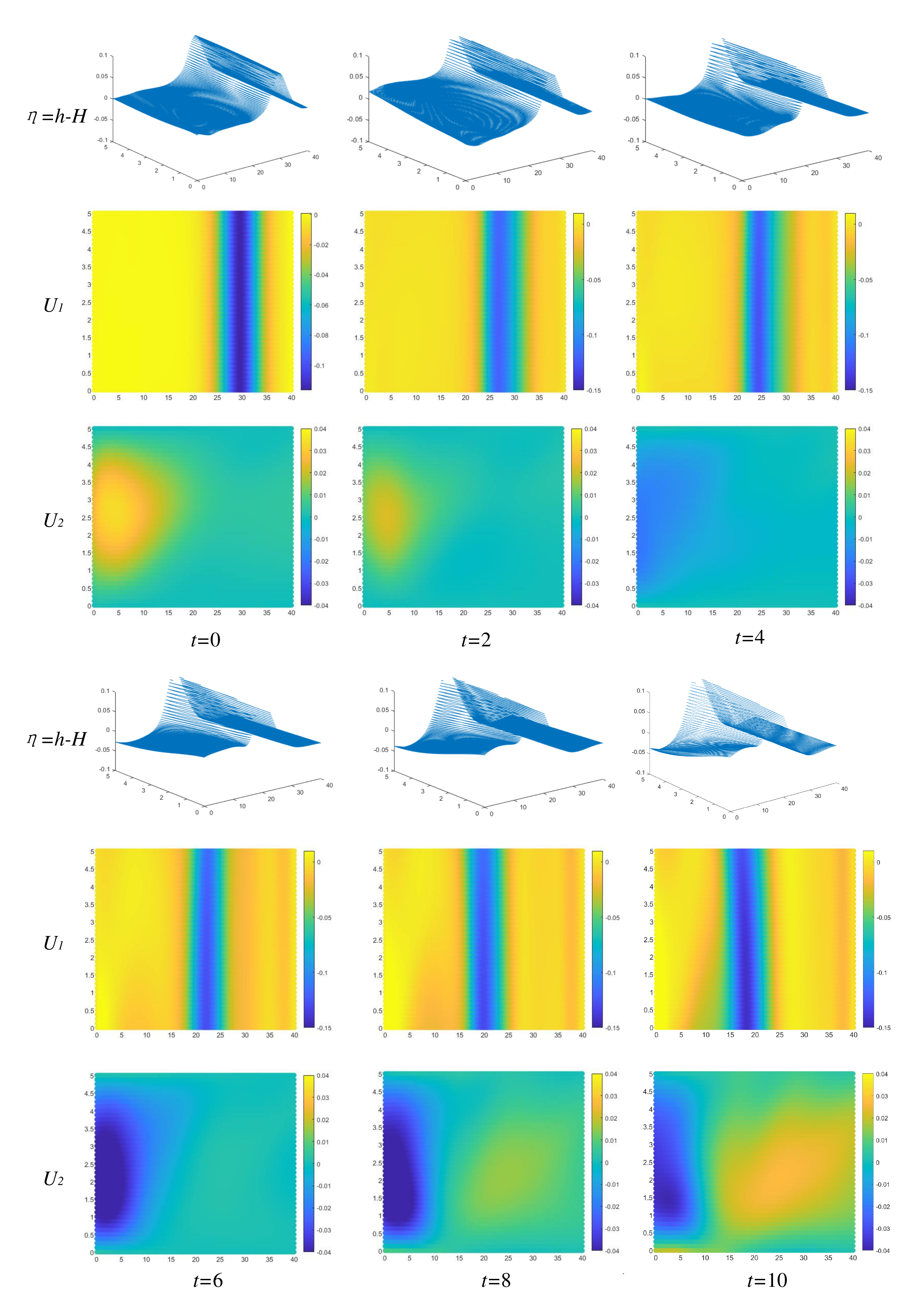}
	\caption{Solitary wave in the inclined bottom flume. Results with incomplete initial conditions. From left to right are the numerical distributions of $U_{i}$ and $\eta$ in the computational domain at time $t=0,2,4,6,8,10$.}
	\label{FIG:4}
\end{figure*}
The initial state of $u_{2}$ is not provided in this case, so the conditions for solving the equations are incomplete. However, the neural network still gives a solution, as shown in Fig. \ref{FIG:4}. Initially, it was thought that when the conditions were incomplete, the result would tend to be a steady solution without a time term. However, it is evident that if the calculation continues, $u_{2}$ will tend to diverge. Therefore, the neural network cannot provide a solution that goes beyond the conventional method and give the correct answer under insufficient conditions. 

\begin{table}[!ht]
	\centering
	\caption{The position of the wave crest in the $x_{1}$-direction. The initial state is known with the initial state of $u_{2}$ is not provided.}\label{tbl6}
	\begin{tabular}{ c|cccccc }
		\hline
		t & 0 & 2 & 4 & 6 & 8 & 10\\
		\hline
		position & 30.0 & 27.3  & 24.8 & 22.5 & 20.5 & 18.0\\
		\hline
	\end{tabular}
\end{table}

Although $u_{2}$ does not tend to $0$ as expected, the solitary wave is still well-described by the neural network. The position of the wave crest in the $x_{1}$-direction over time is shown in Table \ref{tbl6}. The propagation speed is approximately $1.2$, which is not consistent with the expected value. However, the wave height increases during the propagation process, which is consistent with the theoretical results.

\subsection{Solve when the intermediate state is known}
In this section, we will discuss how to obtain the past and the future information when the intermediate time information is known.The boundary conditions are the same as those in Eq. \eqref{eq3.14}. The intermediate state equations are
\begin{equation}\label{eq3.15}
	\begin{aligned}
		&\eta(x_{1},x_{2},5) =a{{\operatorname{sech}}^{2}}\left( \sqrt{3a/4{{h}^{3}}}\times \left( x_{1}-25 \right) \right)\\
		&u_{1}(x_{1},x_{2},5)=-\left( l+1/2a \right)\eta /\left( (x_{1}-5)/30+\eta  \right)\\
		&u_{2}(x_{1},x_{2},5)=0
	\end{aligned}
\end{equation}	

In the previous process, the networks only obtain the future flow field information from the current flow field. We make the networks obtain the past flow field information by adding the time-step items as follows:

Step 1,

Calculate $\Delta U_{i,a}^{*}$ and $\Delta U_{i,c}^{*}$ by Eq. \eqref{eq2.7} and Eq. \eqref{eq3.19}.
\begin{equation}\label{eq3.19}
	\begin{aligned}
		&\Delta U_{i,c}^{*}= \\
		&\Delta t{{\left[ -\frac{\partial \left( {{{\bar{u}}}_{j}}{{U}_{i}} \right)}{\partial {{x}_{j}}}
				+\frac{\Delta t}{2}{{u}_{k}}\frac{\partial }{\partial {{x}_{i}}}\left( \frac{\partial \left( {{{\bar{u}}}_{j}}{{U}_{i}} \right)}{\partial {{x}_{j}}} \right) \right]}^{n}}\\
	\end{aligned}
\end{equation}

Step 2,

Use input $\left( \textbf{x},t-\Delta t \right)^{j}$ to obtain $h_{c}^{j}$. Change $L_{eqns}^{h}$ as
\begin{equation}\label{eq3.20}
	\begin{aligned}
		L_{eqns}^{h}=\sum\limits_{j=1}^{N}{{{\left( {{h}_{a}}^{j}-{{h}_{0}}^{j}-\Delta {{h}_{a}}^{j} \right)}^{2}}/{{\Delta t}^{2}}} 
		+\sum\limits_{j=1}^{N}{{{\left( {{h}_{c}}^{j}-{{h}_{0}}^{j}-\Delta {{h}_{c}}^{j} \right)}^{2}}/{{\Delta t}^{2}}}
	\end{aligned}
\end{equation}	
$\Delta {{h}_{c}}$ is computed as
\begin{equation}\label{eq3.22}
	\Delta {{h}_{c}}=\Delta t{{\left[ \frac{\partial \left( U_{i}^{n} \right)}{\partial {{x}_{j}}}+{{\theta }_{1}}\frac{\partial \Delta U_{i}^{*}}{\partial {{x}_{i}}}-\Delta t{{\theta }_{1}}\frac{{{\partial }^{2}}{{p}^{n-{{\theta }_{2}}}}}{\partial {{x}_{i}}\partial {{x}_{i}}} \right]}}
\end{equation}

Step 3,

Use input $\left( \textbf{x},t-\Delta t \right)^{j}$ to obtain $u_{i,c}^{j}$. Change $L_{eqns}^{u_{i}}$ as
\begin{equation}\label{eq3.23}
	\begin{aligned}
		L_{eqns}^{{{u}_{i}}}=\sum\limits_{j=1}^{N}{{{\left( {{u}_{i,a}}^{j}-\left( {{h}_{0}}{{u}_{i,0}}^{j}+\Delta {{U}_{i,a}}^{j} \right)/{{h}_{a}} \right)}^{2}}/\Delta {{t}^{2}}}\ 
		+\sum\limits_{j=1}^{N}{{{\left( {{u}_{i,c}}^{j}-\left( {{h}_{0}}{{u}_{i,0}}^{j}+\Delta {{U}_{i,c}}^{j} \right)/{{h}_{c}} \right)}^{2}}/\Delta {{t}^{2}}}\
	\end{aligned}
\end{equation}
$\Delta {{U}_{i,c}}$ is computed as
\begin{equation}\label{eq3.24}
	\Delta {{U}_{i,c}}=\Delta U_{i,c}^{*}+\Delta t\frac{\partial p_{i}^{n-{{\theta }_{2}}}}{\partial {{x}_{i}}}+\frac{\Delta {{t}^{2}}}{2}{{\bar{u}}_{k}}\frac{{{\partial }^{2}}{{p}^{n}}}{\partial {{x}_{k}}\partial {{x}_{i}}}
\end{equation}	

\begin{figure*}[htb]
	\centering
	\includegraphics[width=0.7\columnwidth]{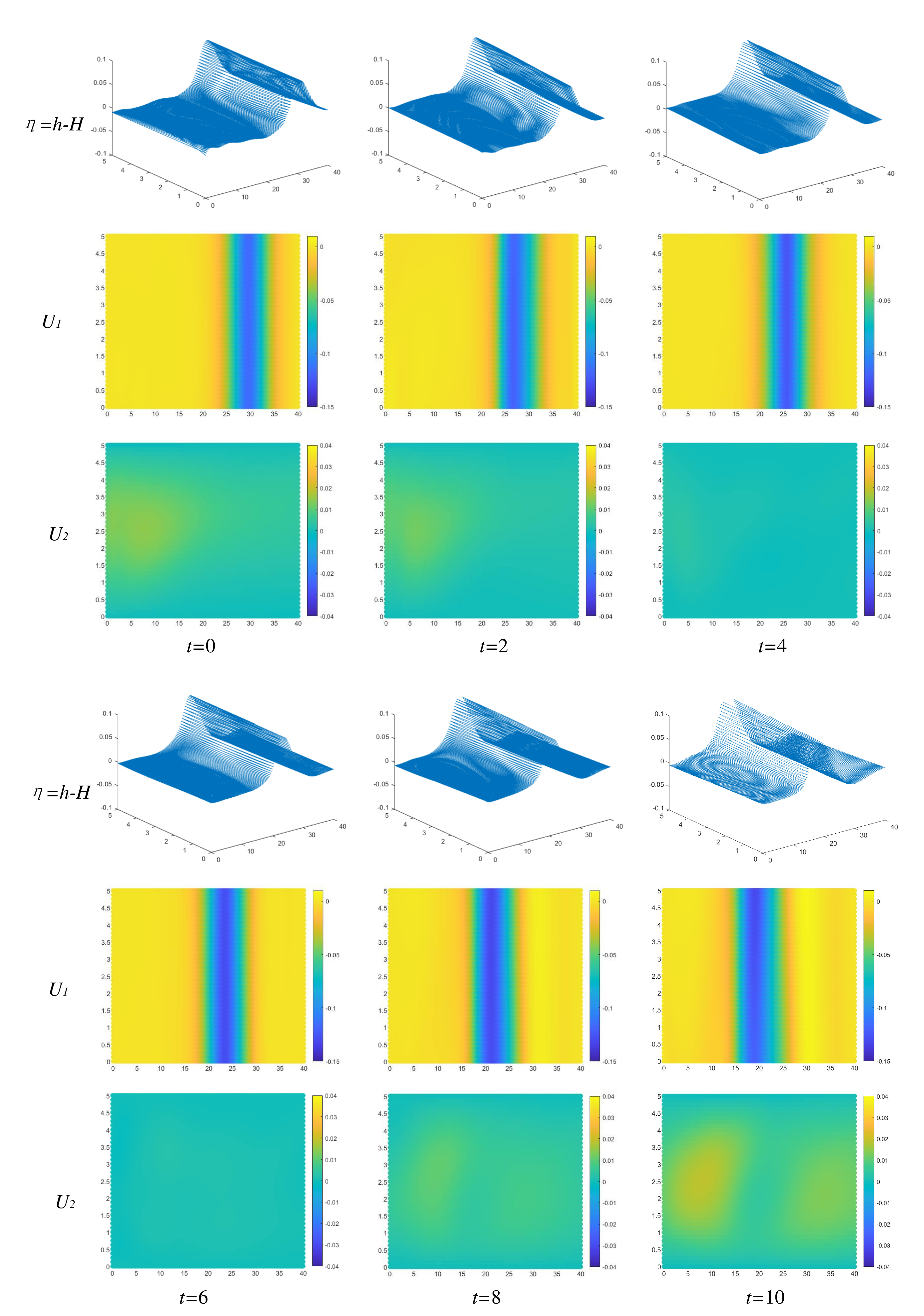}
	\caption{Solitary wave in the inclined bottom flume. Results with complete intermediate conditions. Results with incomplete initial conditions. From left to right are the numerical distributions of $U_{i}$ and $\eta$ in the computational domain at time $t=0,2,4,6,8,10$.}
	\label{FIG:5}
\end{figure*}
When other settings remain unchanged, the modifications mentioned above can achieve better results, as shown in Fig. \ref{FIG:5}. These modifications can improve the stability of the results, but they may also reduce the convergence speed.
The position of the wave crest in the $u_{1}$-direction over time is shown in Table \ref{tbl7}. The propagation speed is approximately $1.1$, which is consistent with the expected value. 

\begin{table}[!htbp]
	\centering
	\caption{The position of the wave crest in the $x_{1}$-direction. The intermediate state is known.}\label{tbl7}
	\begin{tabular}{ c|cccccc }
		\hline
		t & 0 & 2 & 4 & 6 & 8 & 10\\
		\hline
		position & 30.0 & 27.8  & 25.5 & 23.5 & 21.2 & 19.1\\
		\hline
	\end{tabular}
\end{table}

The finite element method inherently implies many boundary conditions, as parameters outside the calculation area do not exist mathematically, and their partial derivatives of all orders are zero. This has both advantages and disadvantages. On the one hand, it does not require setting boundary conditions for the parameter partial derivative term. However, it cannot deal with the reflection of waves at the boundary.

In contrast, the parameters of the neural network outside the calculation area still exist. Therefore, when the wave flows out of the calculation area, it will not be reflected. However, in some cases, more boundary conditions are needed. To address this, the time-step-back items are added in the above method. By considering both the time-step-back and time-step items simultaneously, the method partially achieves the effect that the partial derivatives of the boundary parameters are zero and can replace some boundary conditions.

\subsection{Shallow-water transport}
\label{Sec3.3}
In engineering applications, it is often unnecessary to obtain all the information to guide production. It can be cost-effective to acquire only a part of the information. The separation of output parameters can solve this problem and make it more convenient for further calculations. 

Shallow-water transport problems refer to the study of the transport of fluids with small depths compared to their horizontal extent, such as water in rivers, estuaries, and coastal regions. The transport of such fluids is governed by the shallow-water equations, which are a set of partial differential equations that describe the behavior of such fluids.

Shallow-water transport problems are important in many practical applications, including the management of water resources, flood prediction and control, and the design of hydraulic structures such as dams and levees. In these applications, it is essential to accurately predict the transport of water and other materials such as sediment and pollutants through the shallow-water system.

We can use the flow field information obtained above to calculate the dispersion of some quantities in the shallow-water. The depth-averaged transport equation can be calculated, in which the averaged velocities $u$ have been determined independently. A typical shallow-water transport equation can be written as:
\begin{equation}\label{eq3.25}
	\frac{\partial \left( hT \right)}{\partial t}+\frac{\partial \left( h{{{\bar{u}}}_{i}}T \right)}{\partial {{x}_{i}}}-\frac{\partial }{\partial {{x}_{i}}}\left( hk\frac{\partial T}{\partial {{x}_{i}}} \right)+R=0
\end{equation}	
where $h$ and ${{\bar{u}}}_{i}$ are the previously defined and computed quantities, $k$ is an appropriate diffusion coefficient, and $R$ is a source term. Set $R=0$. 
The application of the CBS method for any scalar transport equation is straightforward, because of the absence of the pressure gradient term. Now a new time integration parameter ${\theta }_{3}$ is introduced for the diffusion term such that $0\le{\theta }_{3}\le1$. We used the method that considers both time-step and time-step-back items to reduce the boundary conditions. The form is:
\begin{equation}\label{eq3.26}
	\begin{aligned}
		 \Delta h{{T}_{a}}=h{{T}^{n+1}}-h{{T}^{n}}=  
		 -\Delta t{{\left[ \frac{\partial \left( h{{{\bar{u}}}_{i}}T_{b} \right)}{\partial {{x}_{i}}}-\frac{\Delta t}{2}{{u}_{k}}\frac{\partial }{\partial {{x}_{k}}}\left( \frac{\partial \left( h{{{\bar{u}}}_{i}}T_{b} \right)}{\partial {{x}_{i}}} \right) \right]}^{n}} 
		+\Delta t\frac{\partial }{\partial {{x}_{i}}}\left( hk\frac{\partial {{T}^{n+{{\theta }_{3}}}}}{\partial {{x}_{i}}} \right)  
	\end{aligned}
\end{equation}	
\begin{equation}\label{eq3.27}
	\begin{aligned}
		 \Delta h{{T}_{c}}=h{{T}^{n-1}}-h{{T}^{n}}= 
		 \Delta t{{\left[ \frac{\partial \left( h{{{\bar{u}}}_{i}}T_{b} \right)}{\partial {{x}_{i}}}+\frac{\Delta t}{2}{{u}_{k}}\frac{\partial }{\partial {{x}_{k}}}\left( \frac{\partial \left( h{{{\bar{u}}}_{i}}T_{b} \right)}{\partial {{x}_{i}}} \right) \right]}^{n}} 
		-\Delta t\frac{\partial }{\partial {{x}_{i}}}\left( hk\frac{\partial {{T}^{n-{{\theta }_{3}}}}}{\partial {{x}_{i}}} \right) 
	\end{aligned}
\end{equation}	
\begin{equation}\label{eq3.28}
	\begin{aligned}
		L_{eqns}^{T}=\sum\limits_{j=1}^{N}{{{\left( {T}_{a}^{j}-\left( {h}_{0}{T}_{0}^{j}+\Delta {{hT}_{a}}^{j} \right)/{{h}_{a}} \right)}^{2}}/\Delta {{t}^{2}}}\ 
		+\sum\limits_{j=1}^{N}{{{\left( {T}_{c}^{j}-\left( {h}_{0}{T}_{0}^{j}+\Delta {{hT}_{c}}^{j} \right)/{{h}_{c}} \right)}^{2}}/\Delta {{t}^{2}}}\
	\end{aligned}
\end{equation}

The computation is carried out in the domain of $20\le x_{1}\le 25$, $0\le x_{2}\le 5 $ and $0\le t\le 10 $.  The sampling intervals are $\Delta x_{1}=0.25$, $\Delta x_{2}=0.25$, and $\Delta t=0.125$, resulting in a total number of sampling points of $N=21\times21\times81=35,721$.

The network use the Adam optimizer with $\alpha^{T}=10$. The initial state equation is 
\begin{equation}\label{eq3.29}
	T\left( x,y,0 \right)=\frac{10}{\sqrt{2\pi }}\exp \left( -\left( {{\left( x-22.5 \right)}^{2}}+{{\left( y-2.5 \right)}^{2}} \right)/2 \right)
\end{equation}
The pseudo code of the calculation process is as follows:
\begin{breakablealgorithm}
	\caption{shallow-water transport problem }
	\begin{algorithmic}[1]
		\State Set $\Delta {{t}}=0.125$
		\State get $T_{ini}$ by Eq. \eqref{eq3.29}
		\For{step in range(1000)} 
		\State input N points $\left( \textbf{x},t \right)$ to obtain $\bar{u}_{i},h$ and $T_{b}$
		\State with torch.no\_grad(): \Comment{Create priori information}
		\State ~~~~${T}_{0}=clone(T_{b})$
		\State ~~~~${T}_{0}(:,t=0)=T_{ini}$
		
		\For{k in range($K_{T}$)}
		\State input N points $\left( \textbf{x},t \right)$, $\left( \textbf{x},t+\Delta t \right)$ and $\left( \textbf{x},t-\Delta t \right)$ to obtain $T_{b}$, $T_{a}$ and $T_{c}$, Respectively
		\State with torch.no\_grad(): 
		\State ~~~~get $\Delta h{{T}_{a}}$ and $\Delta h{{T}_{c}}$ by Eqs. \eqref{eq3.26},\eqref{eq3.27}
		\State get $L_{eqns}^{T}$ by Eqs. \eqref{eq3.28}
		\State $L^{T}.backward()$ and use Adam Optimizer 
		\EndFor
		\EndFor
	\end{algorithmic}
\end{breakablealgorithm}

\begin{figure*}[htb]
	\centering
	\includegraphics[width=1.0\columnwidth]{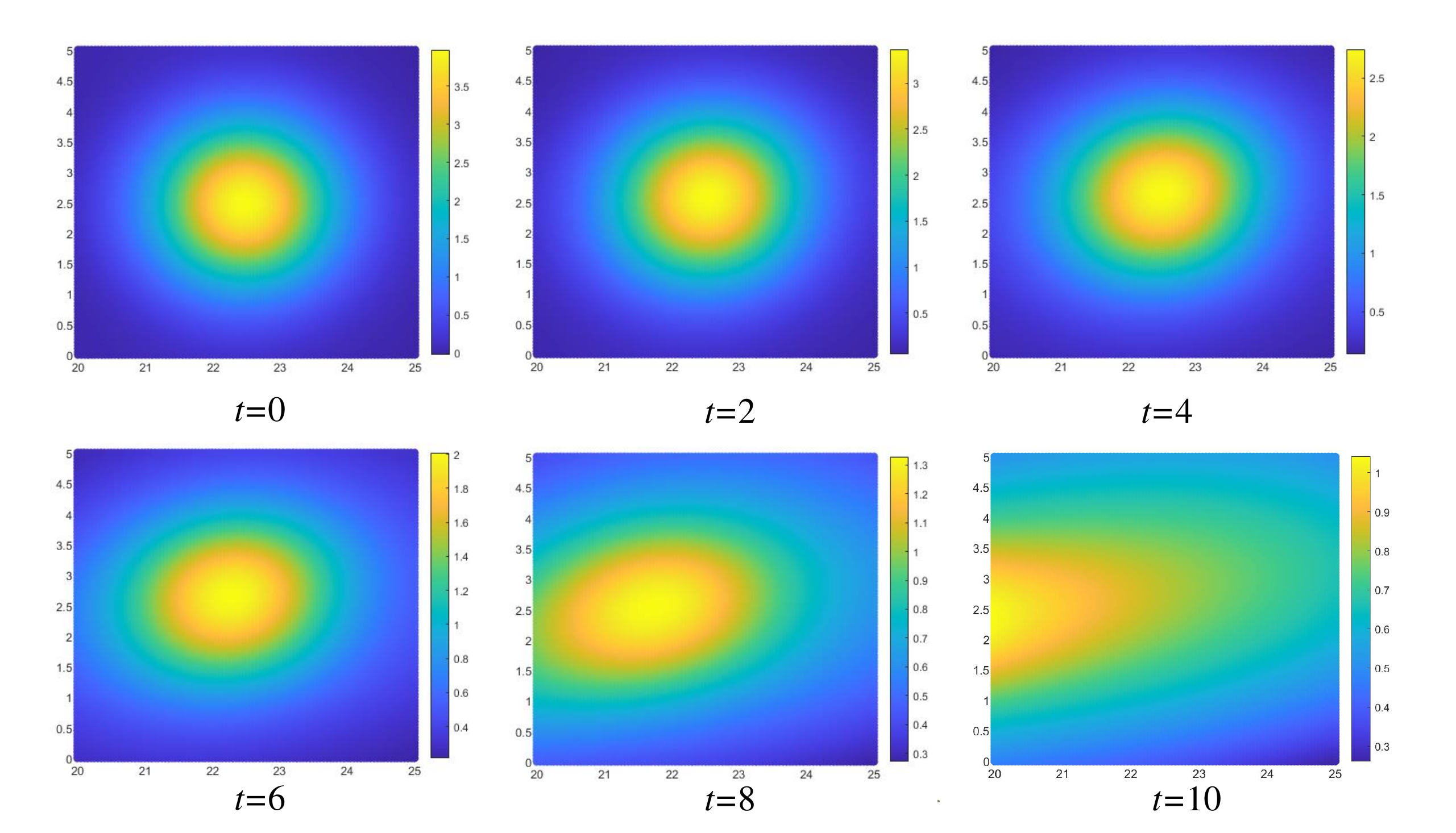}
	\caption{From left to right are temperature fields at time $t=0,2,4,6,8,10$.}
	\label{FIG:6}
\end{figure*}

Stable results can be obtained without boundary conditions, shown in Fig. \ref{FIG:6}.  The temperature field shifts to the right due to the flow field, and the temperature decreases due to diffusion. 

\section{Incompressible flow}
Incompressible flow is a type of fluid flow in which the density of the fluid remains constant, or nearly constant, as it moves through a system. This means that the volume of the fluid is conserved as it flows, and the fluid is not compressed or expanded. Incompressible flow is a simplifying assumption that is often used in the study of fluid mechanics, as it allows for the use of simpler mathematical models and equations.

Incompressible flow has many practical applications, including in the design of pipes, pumps, and other fluid-handling equipment, as well as in the study of aerodynamics. In these applications, the assumption of incompressible flow allows for the use of simpler equations and models that can be used to predict the behavior of the fluid in the system.

As discussed in Section \ref{Sec3}, the shallow water flow equation is a deformation of the N-S equations, and incompressible flow is a special case of the N-S equations. In the context of incompressible flow, we consider the limit as $c\to \infty$, and Eq. \eqref{eq2.1} becomes:
\begin{equation}\label{eq4.1}
	0=-\frac{\partial {{u}_{i}}}{\partial {{x}_{i}}}
\end{equation}
Momentum conservation
\begin{equation}\label{eq4.2}
	\frac{\partial {{u}_{i}}}{\partial t}=-\frac{\partial }{\partial {{x}_{j}}}\left( {{u}_{j}}{{u}_{i}} \right)+\frac{1}{\operatorname{Re}}\frac{{{\partial }^{2}}{{u}_{i}}}{\partial x_{j}^{2}}-\frac{\partial p}{\partial {{x}_{i}}}
\end{equation}
where $Re$ represents the Reynolds number. The rest of the variables are the same as those described before.
Obviously, such simplification leads to the pressure information loss. The change of $p$ is affected by the current flow field. Water is generally considered as an incompressible flow, and its sound velocity is  $c=1480m/s$. In the finite element method, there are two ways to deal with it. One is to retain the $p$ item in the mass formula, so the time accuracy will become extremely high. If the mesh size is  $\Delta x=0.1m$, then the time step should meet the requirement $\Delta t=0.1/1480s$. The other is to use the implicit scheme like CBS. The three essential steps of the CBS scheme can be written in its semi-discrete form as
\begin{figure*}[htbp]
	\centering
	\includegraphics[width=0.70\columnwidth]{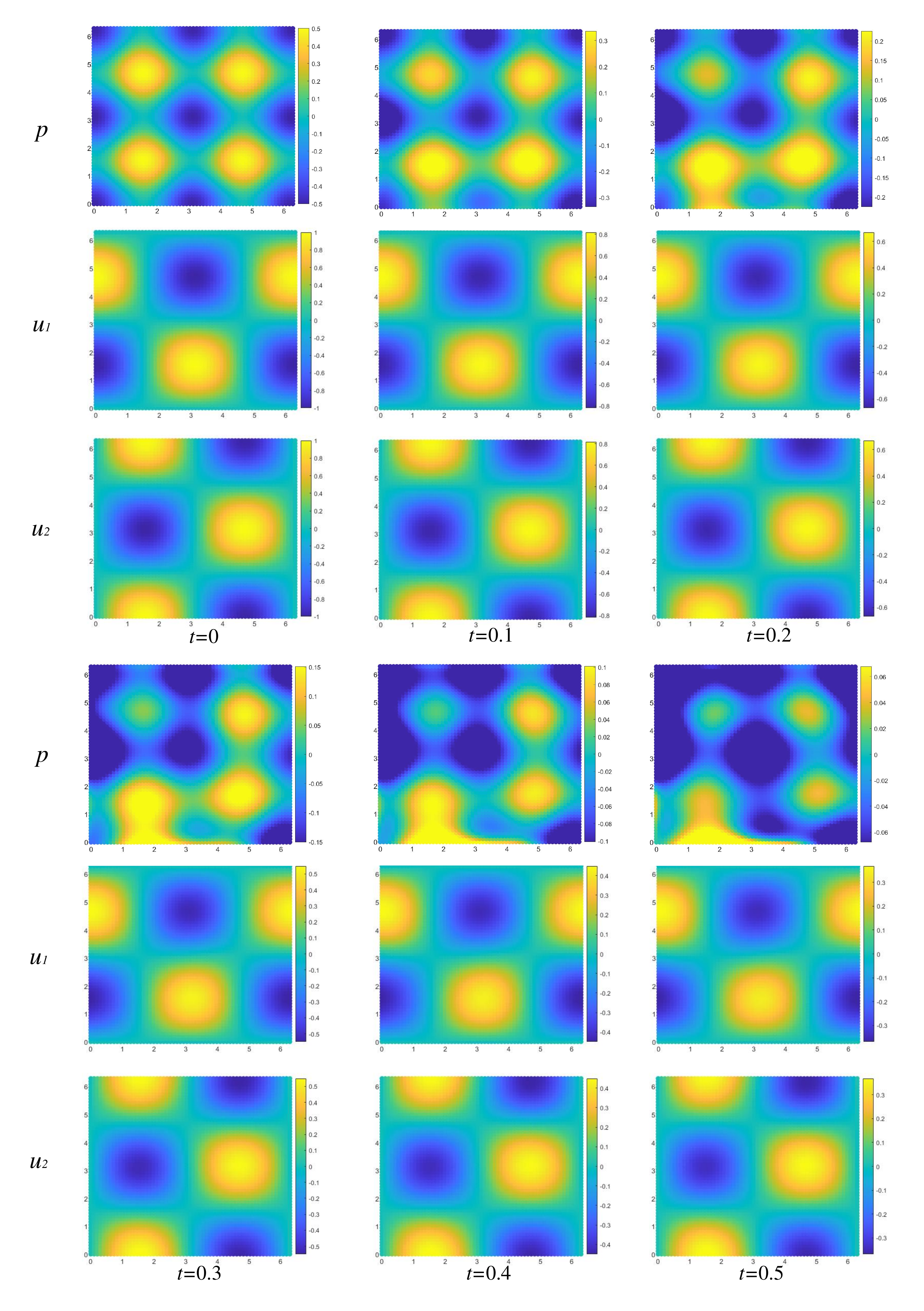}
	\caption{The spatial distribution of $p$, $u_{1}$ and $u_{2}$ from top to bottom. From left to right is their change with time. Boundary conditions without $p$.}
	\label{FIG:8}
\end{figure*}

Step 1,
\begin{equation}\label{eq4.3}
	\begin{aligned}
		\Delta u_{i,a}^{*}= 
		\Delta t{{\left[ -\frac{\partial \left( {{u}_{j}}{{u}_{i}} \right)}{\partial {{x}_{j}}}+\frac{\Delta t}{2}{{u}_{k}}\frac{\partial }{\partial {{x}_{i}}}\left( \frac{\partial \left( {{u}_{j}}{{u}_{i}} \right)}{\partial {{x}_{j}}} \right)+\frac{1}{\operatorname{Re}}\frac{{{\partial }^{2}}{{u}_{i}}}{\partial x_{j}^{2}} \right]}^{n}}
	\end{aligned}
\end{equation}

Step 2,
\begin{equation}\label{eq4.4}
	\Delta _{a}=\Delta t\left[ \frac{\partial \left( u_{i}^{n} \right)}{\partial {{x}_{j}}}+{{\theta }_{1}}\frac{\partial \Delta u_{i,a}^{*}}{\partial {{x}_{i}}}+\Delta t{{\theta }_{1}}\frac{{{\partial }^{2}}{{p}^{n+{{\theta }_{2}}}}}{\partial {{x}_{i}}\partial {{x}_{i}}} \right]
\end{equation}
\begin{equation}\label{eq4.13}
	L_{Prior}^{p}=\sum\limits_{j=1}^{N}{{{\left( {{p}_{0}^{j}}-{{p}_{b}^{j}} \right)}^{2}}/\Delta{{t}^{2}}}
\end{equation}
\begin{equation}\label{eq4.14}
	L_{eqns}^{p}=\sum\limits_{j=1}^{N}{{{\left( \Delta _{a} \right)}^{2}}/\Delta{{t}^{2}}}
\end{equation}

Step 3,
\begin{equation}\label{eq4.5}
	\Delta {{u}_{i,a}}=\Delta u_{i,a}^{*}-\Delta t\frac{\partial p_{i}^{n+{{\theta }_{2}}}}{\partial {{x}_{i}}}+\frac{\Delta {{t}^{2}}}{2}{{u}_{k}}\frac{{{\partial }^{2}}{{p}^{n}}}{\partial {{x}_{k}}\partial {{x}_{i}}}
\end{equation}
\begin{equation}\label{eq4.16}
	L_{Prior}^{u_{i}}=\sum\limits_{j=1}^{N}{{{\left( {{u_{i,0}}^{j}}-{{u_{i,b}}^{j}} \right)}^{2}}/\Delta{{t}^{2}}}
\end{equation}
\begin{equation}\label{eq4.17}
	L_{eqns}^{{{u}_{i}}}=\sum\limits_{j=1}^{N}{{{\left( {u}_{i,a}^{j}-u_{i,0}^{j}-\Delta {{u}_{i,a}}^{j} \right)}^{2}}/\Delta {{t}^{2}}}
\end{equation}

After Step 1, the calculation graph generated in this step is destroyed and only the value of $\Delta u_{i}^{*}$ is retained. This step does not involve backpropagation. After Step 2, only the calculation graphs that generate $\frac{{{\partial }^{2}}{p}}{\partial {{x}_{i}}\partial {{x}_{i}}}$ and $p$ are retained. After step 3, only the calculation graph of $u_{i}$ is retained. Only parameters related to the retained computational graph can participate in backpropagation.
\subsection{Boundary conditions}
We consider 2D Taylor's decaying vortices to check the method. When $Re=1$, an exact 2D solution to Eqs. \eqref{eq4.1}-\eqref{eq4.2} given by \citet{ref17} and \citet{ref31} is as follows:
\begin{equation}\label{eq4.6}
	\begin{aligned}
		& u_{1}\left( x_{1},x_{2},t \right)=-\cos \left( x_{1} \right)\sin \left( x_{2} \right){{e}^{-2t}} \\ 
		& u_{2}\left( x_{1},x_{2},t \right)=\sin \left( x_{1} \right)\cos \left( x_{2} \right){{e}^{-2t}} \\ 
		& p\left( x_{1},x_{2},t \right)=-0.25\left( \cos \left( 2x_{1} \right)+\cos \left( 2x_{2} \right) \right){{e}^{-4t}} \\ 
	\end{aligned}
\end{equation}

The computation is carried out in the domain of $0\le x_{1}\le 2\pi$, $0\le x_{2}\le 2\pi$ and $0\le t\le 0.5 $. The sampling intervals are $\Delta x_{1}=0.05\pi$, $\Delta x_{2}=0.05\pi$, and $\Delta t=0.025$, resulting in a total number of sampling points of $N=41\times41\times21=35,301$.
All three networks use the Adam optimizer and $\alpha^{p}=0.01$, $\alpha^{u_{1}}=\alpha^{u_{2}}=0.1$. The initial state equations and boundary conditions are
\begin{equation}\label{eq4.7}
	\begin{aligned}
		& u_{1}\left( x_{1},x_{2},0 \right)=-\cos \left( x_{1} \right)\sin \left( x_{2} \right) \\ 
		& u_{2}\left( x_{1},x_{2},0 \right)=\sin \left( x_{1} \right)\cos \left( x_{2} \right) \\ 
		& p\left( x_{1},x_{2},0 \right)=-0.25\left( \cos \left( 2x_{1} \right)+\cos \left( 2x_{2} \right) \right) \\ 
	\end{aligned}
\end{equation}	
\begin{equation}\label{eq4.8}
	\left\{ \begin{aligned}
		& u_{1}\left( 0,{{x}_{2}},t \right)= u_{1}\left( 2\pi,{{x}_{2}},t \right)=-\sin (x_{2}) \\ 
		& u_{1}\left({{x}_{2}},0,t \right)= u_{1}\left( {{x}_{1}},2\pi,t \right)=0\\ 
		& u_{2}\left( 0,{{x}_{2}},t \right)= u_{2}\left( 2\pi,{{x}_{2}},t \right)=0 \\ 
		& u_{2}\left( {{x}_{1}},0,t \right)= u_{2}\left( {{x}_{1}},2\pi,t \right)=\sin (x_{1}) \\ 
	\end{aligned} \right.
\end{equation}	

The results are presented in Fig. \ref{FIG:8}. It can be observed that the results for $u_{1}$ and $u_{2}$ are more consistent with the exact solution, while $p$ deviates significantly from the exact solution. It is evident that $p$ exhibits significant leakage at the boundary.

In order to get the correct results, additional boundary conditions for pressure need to be added as
\begin{equation}\label{eq4.9}
	\left\{ \begin{aligned}
		& p\left( 0,{{x}_{2}},t \right)= p\left( 2\pi,{{x}_{2}},t \right)=-0.25(1+\cos(2x_{2})) \\ 
		& p\left( {{x}_{1}},0,t \right)= p\left( {{x}_{1}},2\pi,t \right)=-0.25(1+\cos(2x_{1})) \\ 
	\end{aligned} \right.
\end{equation}	

\begin{figure}[htbp]
	\centering
	\includegraphics[width=0.8\columnwidth]{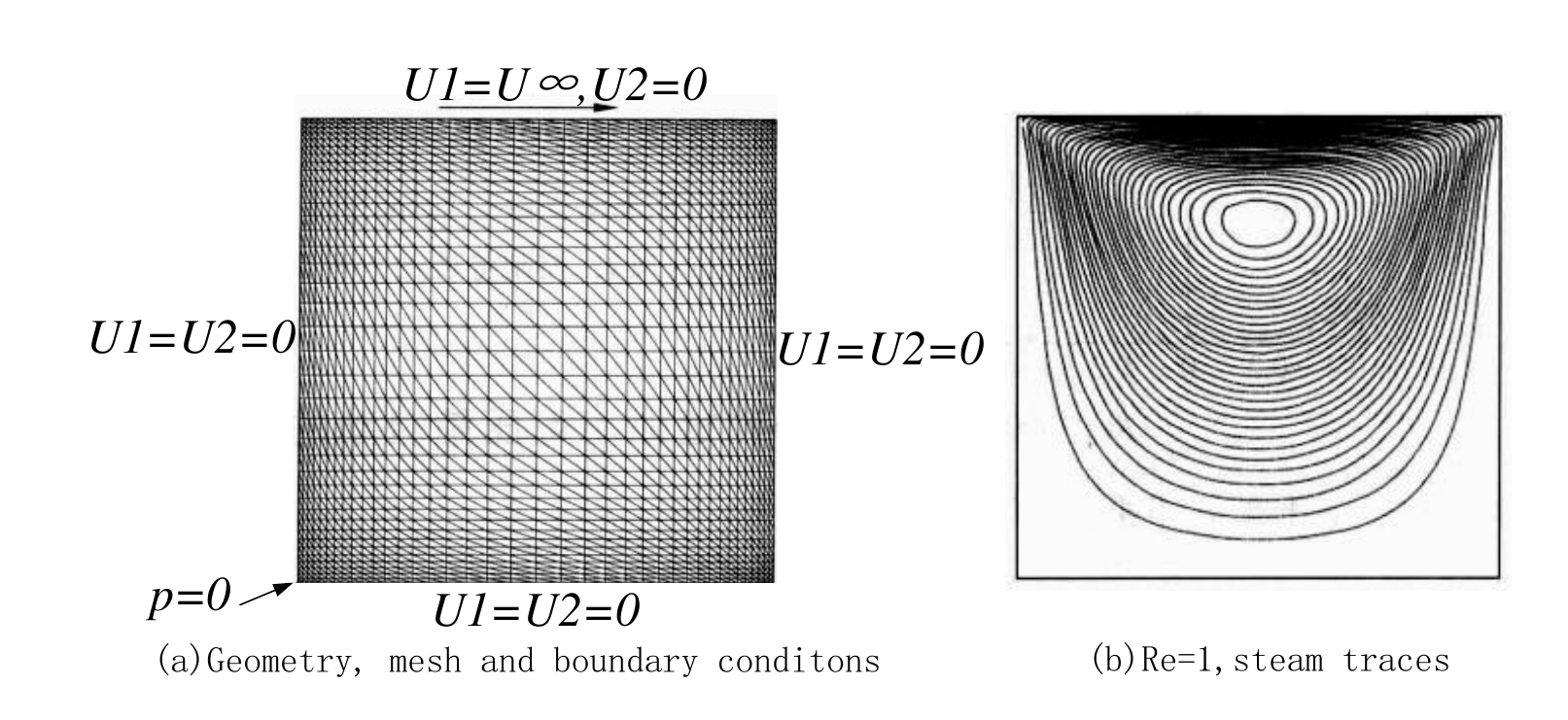}
	\caption{Inconpressible flow in a lid-driven cavity. (a) Geometry, mesh and boundary conditions; (b) Re=1, steam traces. }
	\label{FIG:7}
\end{figure}
A classic problem, namely a closed cavity driven by the motion of a lid \citep{ref18}, highlights the need for additional boundary conditions, as illustrated in Fig. \ref{FIG:7}. The results demonstrate that the velocity component parallel to the wall has a significant impact on the overall flow field, emphasizing the role of the viscous term. In the finite element method, there are no parameters outside the computational domain, and the values on the boundaries are solely influenced by the parameters within the domain. However, in neural networks, parameters outside the computational domain can continuously affect the values on the boundaries. This is analogous to applying random parallel wall velocity components to the four walls of a cavity. Therefore, neural networks require more boundary conditions to constrain the results, whereas the finite element method can omit certain boundary conditions when dealing with some symmetric problems.

The pseudo code of the calculation process is as follows:
\begin{breakablealgorithm}
	\caption{Incompressible N-S equations problem }
	\begin{algorithmic}[1]
		\State Set $\Delta {{t}}=0.005$
		\State get $p_{ini},u_{i,ini}$ by Eq. \eqref{eq3.12}
		\For{step in range(1000)} 
		
		\State //Step 1 
		\State input N points $\left( \textbf{x},t \right)$ to obtain $\bar{u}_{i,b},h_{b}$
		\State with torch.no\_grad(): \Comment{Create priori information}
		\State ~~~~$u_{i,0},p_{0}=clone(u_{i,b},p_{b})$
		\State ~~~~$p_{0}(:,t=0),u_{i,0}(:,t=0)=p_{ini},u_{i,ini}$
		\State ~~~~add boundary conditions by Eqs. \eqref{eq4.8},\eqref{eq4.9}
		\State get $\Delta u_{i,a}^{*}$ by Eq. \eqref{eq4.3}.     
		
		\State //Step 2 
		\State with torch.no\_grad(): \Comment{Prepare for $\Delta {{p}_{a}}$}
		\State ~~~~get $\frac{\partial \left( u_{i}^{n} \right)}{\partial {{x}_{j}}},\frac{\partial \Delta u_{i}^{*}}{\partial {{x}_{i}}}$ by AD
		\For{$k$ in range($K_{p}$)}
		\State input N points $\left( \textbf{x},t \right)$ and $\left( \textbf{x},t+\Delta t \right)$ to obtain $p_{b}$ and $p_{a}$, Respectively 
		\State get $\frac{{{\partial }^{2}}{{p}^{n+{{\theta }_{2}}}}}{\partial {{x}_{i}}\partial {{x}_{i}}}$ \Comment{Gradient is mandatory}
		\State get $\Delta _{a}$ by Eq. \eqref{eq4.4}
		\State get $L^{p}$ by Eqs. \eqref{eq4.13},\eqref{eq4.14}
		\State $L^{p}.backward()$ and use Adam Optimizer 
		\EndFor
		
		\State //Step 3
		\State with torch.no\_grad(): 
		\State ~~~~input N points $\left( \textbf{x},t+\Delta t \right)$ to obtain $p_{a}$
		\State ~~~~get $\frac{\partial p_{i,a}}{\partial {{x}_{i}}}$
		\State ~~~~get $\Delta {{u}_{i,a}}$ by Eq. \eqref{eq4.5}
		\For{$k$ in range($K_{u_{i}}$)}
		\State input N points $\left( \textbf{x},t \right)$ and $\left( \textbf{x},t+\Delta t \right)$ to obtain $u_{i,b}$ and $u_{i,a}$, Respectively
		\State get $L^{u_{i}}$ by Eqs. \eqref{eq4.16},\eqref{eq4.17}
		\State $L^{u_{i}}.backward()$ and use Adam Optimizer 
		\EndFor
		\EndFor
	\end{algorithmic}
\end{breakablealgorithm}

\begin{figure*}[htbp]
	\centering
	\includegraphics[width=0.8\columnwidth]{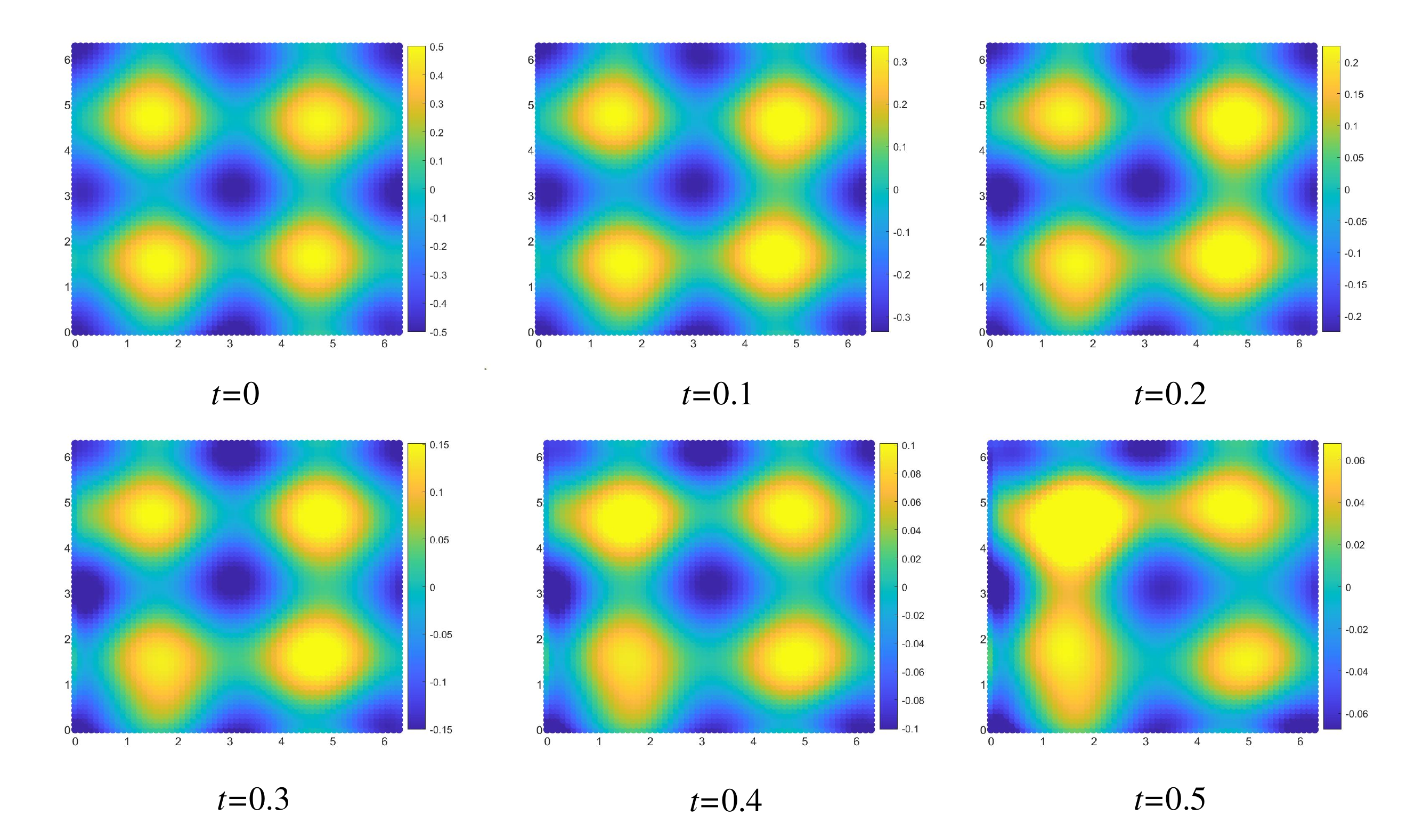}
	\caption{The spatial distribution of $p$. From left to right is their change with time. Boundary conditions with $p$. $\Delta t=0.0125$}
	\label{FIG:9}
\end{figure*}
\begin{figure*}[h]
	\centering
	\includegraphics[width=0.8\columnwidth]{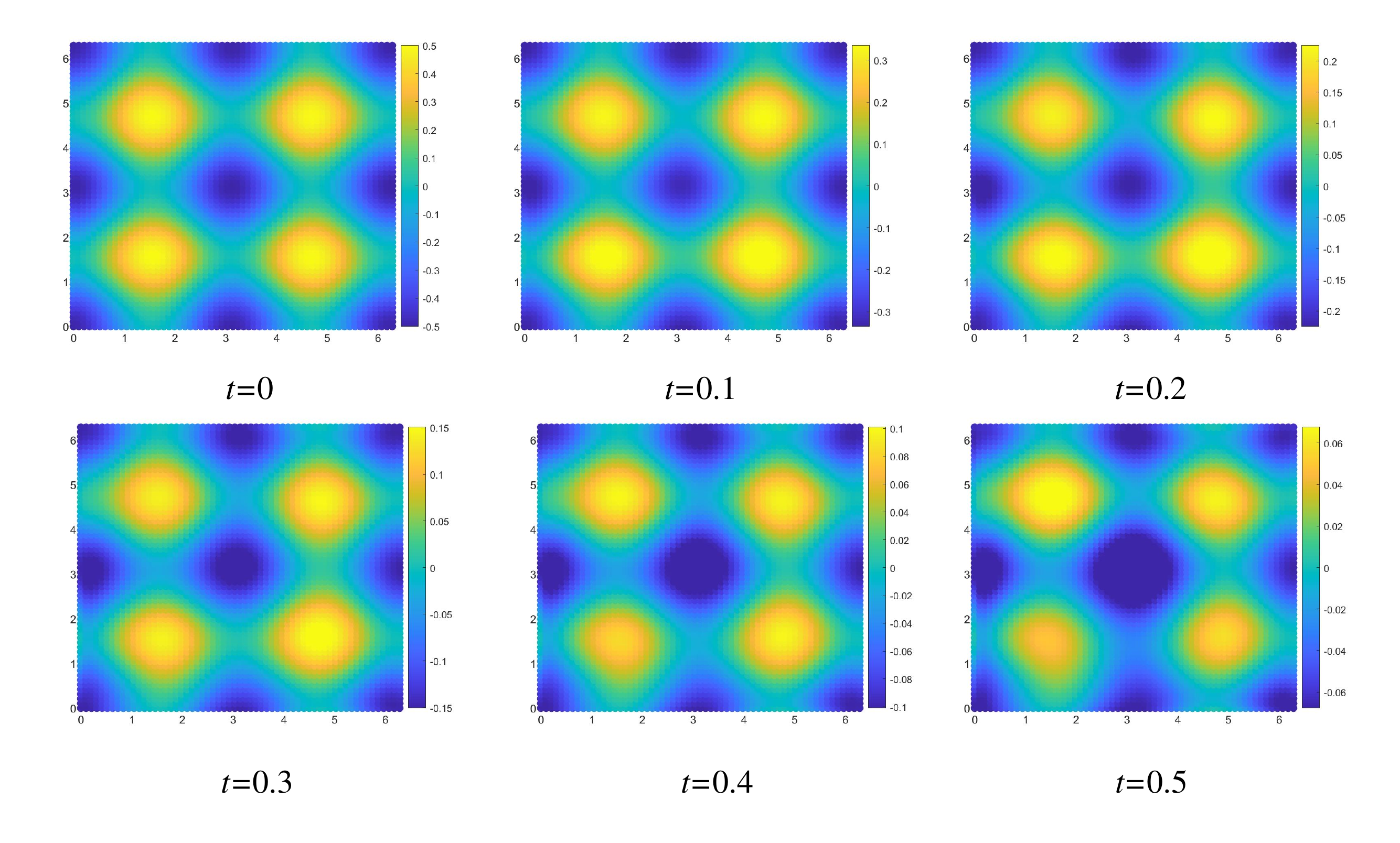}
	\caption{The spatial distribution of $p$. From left to right is their change with time. Boundary conditions with $p$. $\Delta t=0.005$}
	\label{FIG:10}
\end{figure*}
Set $\alpha^{p}=0.01$, $\alpha^{u_{1}}=\alpha^{u_{2}}=10$, while keeping all other conditions unchanged, the results obtained using Eq. \eqref{eq4.9} are presented in Fig. \ref{FIG:9}. The $L_2$ norm, which can be utilized to evaluate the accuracy of the results, is calculated for $p$, $u_{1}$, and $u_{2}$, resulting in values of 0.17, 0.017, and 0.015, respectively. Since $p$ is a high-frequency parameter relative to $u_{1}$ and $u_{2}$, the F-principle is well reflected here. In this case, the simplest approach to address the issue is by reducing the time step $\Delta t$. By setting $\Delta t=0.005$, while keeping all other conditions unchanged, the $L_2$ norm of $p$, $u_{1}$, and $u_{2}$ can be reduced to 0.13, 0.0069, and 0.0071, respectively, as shown in Fig. \ref{FIG:10}.

\subsection{Factors that affect accuracy}
Table \ref{tbl2} presents a detailed systematic study that quantifies the effect of different time-step sizes $\Delta t$. As shown in the table, it can be observed that decreasing the time-steps $\Delta t$ leads to a reduction in the relative final prediction errors. By keeping $\Delta x_{1}=0.05\pi$, $\Delta x_{2}=0.05\pi$, and $\Delta t=0.005$ constant, we vary the total calculation time $t_{final}$ and present the resulting relative final prediction errors for $p$, $u_{1}$, and $u_{2}$ in Table \ref{tbl3}. It is noteworthy that the errors increase with the increase of $t_{final}$. It is difficult to determine whether this is due to insufficient network storage space or the accumulation of time-stepping errors. The impact of network architecture on prediction accuracy is discussed in Table \ref{tbl4}. It is found that only when the network architecture is less than 2 hidden layers with 32 neurons per layer, will the prediction accuracy of the network be significantly decreased.

\begin{table}[!htbp]
	\centering
	\caption{Relative final prediction errors of $p$, $u_{1}$ and $u_{2}$ measured in the $L_{2}$ norm for different time-steps $\Delta t$. Sampling interval in space is fixed to $\Delta x_{1}=0.05\pi$ and $\Delta x_{2}=0.05\pi$. The network architecture is 3 hidden layers with 64 neurons per layer.}\label{tbl2}
	\begin{tabular}{ c|ccccc }
		\hline
		\diagbox{Parameters}{$\Delta t$} & 0.1 & 0.05 & 0.025 & 0.0125 & 0.005 \\
		\hline
		$p$ & 4.5e-1 & 3.3e-1 & 2.5e-1 & 1.7e-1 & 1.3e-1\\
		$u_{1}$ & 2.7e-2 & 2.2e-2 & 2.0e-2 & 1.7e-2 & 6.9e-3\\
		$u_{2}$ & 2.6e-2 & 2.4e-2 & 2.0e-2 & 1.5e-2 & 7.1e-3\\
		\hline
	\end{tabular}
\end{table}

\begin{table}[!ht]
	\centering
	\caption{Relative final prediction errors of $p$, $u_{1}$ and $u_{2}$ measured in the $L_{2}$ norm for different total time $t_{final}$. Sampling interval is fixed to $\Delta x_{1}=0.05\pi$, $\Delta x_{2}=0.05\pi$ and $\Delta t=0.005$. The network architecture is 3 hidden layers with 64 neurons per layer.}\label{tbl3}
	\begin{tabular}{ c|ccccc }
		\hline
		\diagbox{Parameters}{$t_{final}$} & 0.1 & 0.2 & 0.3 & 0.4 & 0.5 \\
		\hline
		$p$ & 6.4e-2 & 9.4e-2 & 1.1 e-1& 1.2e-1 &  1.3e-1\\
		$u_{1}$ & 2.0e-3 & 3.0e-3 & 4.2e-3 & 5.5e-3 & 6.9e-3\\
		$u_{2}$ & 2.1e-3 & 3.4e-3 & 4.8e-3 & 6.3e-3 & 7.1e-3\\
		\hline
	\end{tabular}
\end{table}

\begin{table}[!ht]
	\centering
	\caption{Relative final prediction errors of $p$, $u_{1}$ and $u_{2}$ measured in the $L_{2}$ norm for different number of hidden layers and neurons per layer. Sampling interval is fixed to $\Delta x_{1}=0.05\pi$, $\Delta x_{2}=0.05\pi$ and $\Delta t=0.005$.}\label{tbl4}
	\begin{tabular}[c]{ c|cccc }
		\hline
		\multicolumn{5}{|c|}{Relative final prediction errors of $p$}\\
		\cline{1-5}
		\diagbox{Layers}{Neurons} & 16 & 32 & 64 & 128 \\
		\cline{1-5}
		2 & 3.0e-1 & 2.1e-1 & 1.5e-1 & 1.4e-1 \\
		3 & 1.7e-1 & 1.4e-1 & 1.3e-1 & 1.3e-1 \\
		4 & 1.4e-1 & 1.3e-1 & 1.3e-1 & 1.3e-1 \\
		\cline{1-5}
	\end{tabular}
	\begin{tabular}{ c|cccc }
		\hline
		\multicolumn{5}{|c|}{Relative final prediction errors of $u_{1}$}\\
		\cline{1-5}
		\diagbox{Layers}{Neurons} & 16 & 32 & 64 & 128 \\
		\cline{1-5}
		2 & 2.9e-2 & 2.0e-2 & 8.5e-3 & 9.5e-3 \\
		3 & 8.7e-3 & 8.4e-3 & 6.7e-3 & 7.6e-3\\
		4 & 7.0e-3 & 8.1e-3 & 7.2e-3 & 6.9e-3 \\
		\cline{1-5}
	\end{tabular}
	\begin{tabular}{ c|cccc }
		\hline
		\multicolumn{5}{|c|}{Relative final prediction errors of $u_{2}$}\\
		\cline{1-5}
		\diagbox{Layers}{Neurons} & 16 & 32 & 64 & 128 \\
		\cline{1-5}
		2 & 2.1e-2 & 1.7e-2 & 8.3e-3 & 8.7e-3 \\
		3 & 8.9e-3 & 8.0e-3 & 7.2e-3 & 7.5e-3 \\
		4 & 7.7e-3 & 8.0e-3 & 7.0e-3 & 7.1e-3 \\
		\cline{1-5}
	\end{tabular}
\end{table}

\begin{figure*}[htb]
	\centering
	\begin{minipage}[t]{.45\linewidth}
		\centering
		\includegraphics[width=\textwidth]{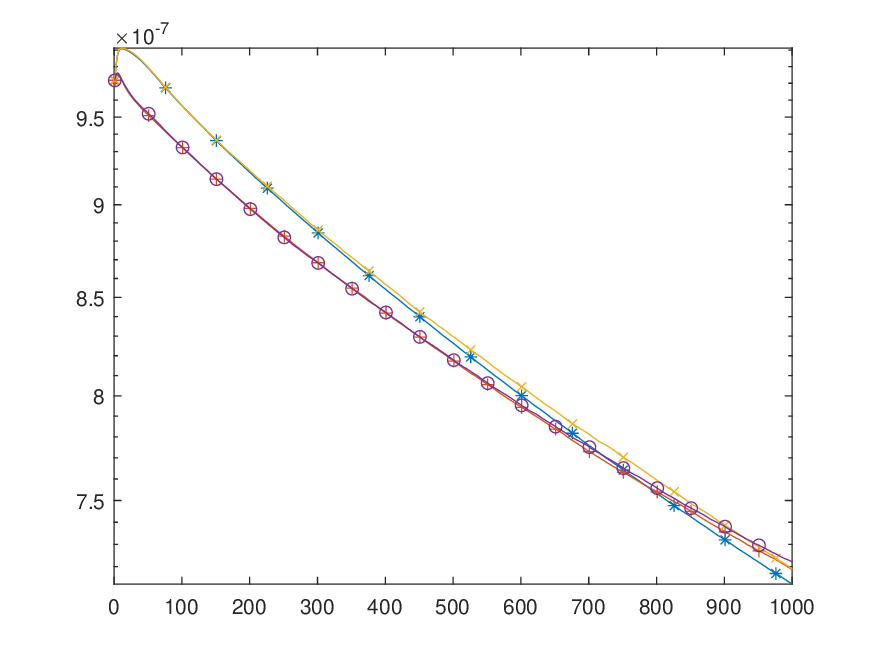}
		\centerline{$(a)$ $L^{P}$}
	\end{minipage}
	\begin{minipage}[t]{.45\linewidth}
		\centering
		\includegraphics[width=\textwidth]{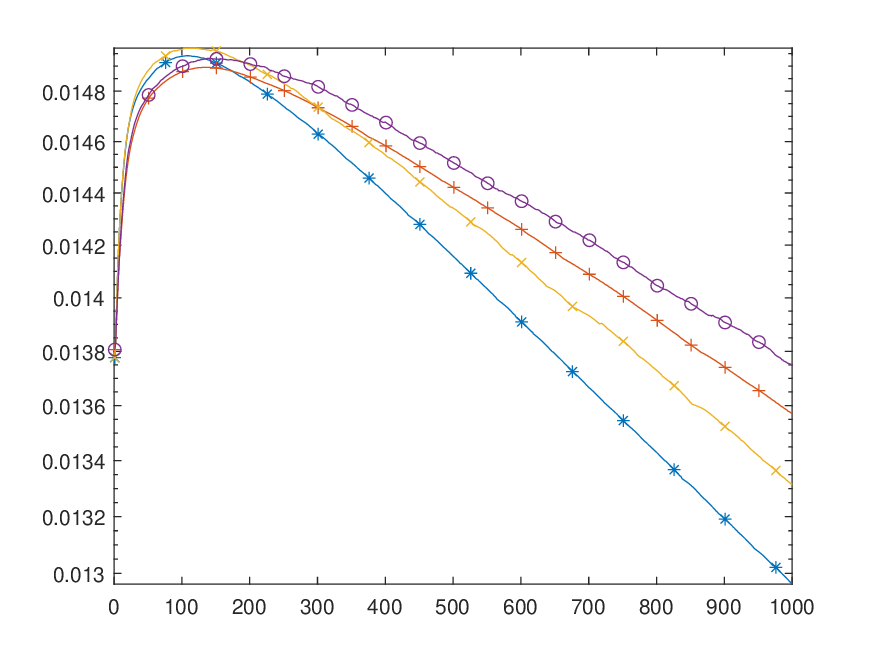}
		\centerline{$(b)$ $L^{u_{1}}$}
	\end{minipage}
	\begin{minipage}[t]{.45\linewidth}
		\centering
		\includegraphics[width=\textwidth]{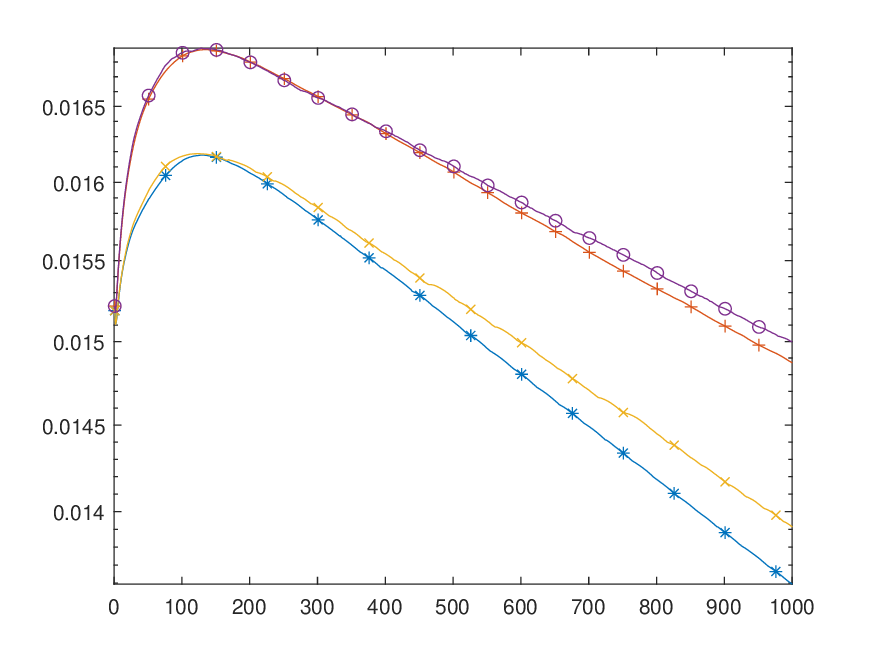}
		\centerline{$(b)$ $L^{u_{2}}$}
	\end{minipage}
	\caption{The Loss function of the parameter $p$ and $u_{i}$ changes with the different number of variables involved in gradient backpropagation. $*$ represents the original algorithm. $+$, x and $o$ represent $\frac{{\partial }{p}}{\partial {{x}_{i}}}$, $\frac{\partial u_{i}}{\partial {{x}_{i}}}$ and $\frac{{\partial }{p}}{\partial {{x}_{i}}}$ add $\frac{\partial u_{i}}{\partial {{x}_{i}}}$, respectively. Sampling inccerval is fixed to $\Delta x_{1}=0.05\pi$, $\Delta x_{2}=0.05\pi$ and $\Delta t=0.005$. $t_{final}=0.2$. The network architecture is 3 hidden layers with 64 neurons per layer.}
	\label{FIG:12}
\end{figure*}

To further investigate the impact of blocking variables from participating in gradient backpropagation on accuracy, we examine the effect of increasing the number of variables involved in gradient backpropagation and compare it with the original algorithm. In the original algorithm, only one variable participated in gradient backpropagation during each step, namely $\frac{{{\partial }^{2}}{p}}{\partial {{x}{i}}\partial {{x}{i}}}$ and $u_{i}$. We then introduce two additional variables, namely $\frac{{\partial }{p}}{\partial {{x}{i}}}$ and $\frac{\partial u{i}}{\partial {{x}_{i}}}$, and compare the results with the original algorithm. The term "NULL" denotes the original algorithm.

We only consider first-order variables to participate in gradient backpropagation, as involving second-order variables in gradient backpropagation would significantly increase memory usage and reduce computational efficiency.

\begin{table*}[!ht]
	\centering
	\caption{Relative final prediction errors of $p$, $u_{1}$ and $u_{2}$ measured in the $L_{2}$ norm for different additional parameters involved in gradient backpropagation. Sampling inccerval is fixed to $\Delta x_{1}=0.05\pi$, $\Delta x_{2}=0.05\pi$ and $\Delta t=0.005$. $t_{final}=0.2$. The network architecture is 3 hidden layers with 64 neurons per layer.}\label{tbl5}
	\begin{tabular}{ ccccc }
		\hline
		&\multicolumn{4}{@{}c@{}}{Additional variables} \\\cmidrule{2-5}
		Parameters & NULL & $\frac{{\partial }{p}}{\partial {{x}_{i}}}$ & $\frac{\partial u_{i}}{\partial {{x}_{i}}}$ &  $\frac{{\partial }{p}}{\partial {{x}_{i}}}$ and $\frac{\partial u_{i}}{\partial {{x}_{i}}}$ \\
		\hline
		$p$     & 2.12e-1 & 2.13e-1 & 2.10e-1 & 2.12e-1\\
		$u_{1}$ & 7.07e-3 & 7.06e-3 & 7.12e-3 & 7.06e-3\\
		$u_{2}$ & 7.36e-3 & 7.44e-3 & 7.42e-2 & 7.46e-2\\
		\hline
	\end{tabular}
\end{table*}

The effect of the number of variables involved in gradient backpropagation is illustrated in Fig. \ref{FIG:12} and summarized in Table \ref{tbl5}. From the decline curve of the loss function, it can be observed that the decline speed of the loss function decreases with an increase in the number of variables involved in gradient backpropagation. However, in terms of final accuracy, the additional variables did not have a significant impact on the accuracy of the results. It is not entirely reliable to infer the solution accuracy based solely on the loss function.

Overall, the key parameters that control the performance of our method are the time-step $\Delta t$, which is consistent with the conclusion obtained by \citet{ref30} using the Runge-Kutta method.

\section{Conclusions}
This paper presents a PINN method based on CBS that can effectively solve shallow-water  equations, making it suitable for ocean flow field estimation. By solving incompressible N-S equations with exact solutions, we prove the correctness and universality of this method. Compared to other methods, our approach is computationally efficient as it does not consider the weights between output parameters, and not all partial derivatives are involved in gradient backpropagation. We investigate the minimum number of conditions required to ensure the reliability and stability of the network output. Our method can provide reliable results even with minimal data volumes. We also discuss the circumstances in which this method requires more stringent conditions than the finite element method. This is mainly due to the partial derivative term in the PINN not being zero at the boundary, which can lead to inaccurate predictions if not taken into consideration.

We compared the effects of network structure, time-step, and gradient backpropagation on the results and found that the network structure is not the primary factor affecting the accuracy, but rather the method itself. In our proposed method, the time-step plays a crucial role in determining the accuracy. Like the finite element method, the neural network is a spatiotemporal approximation method that requires corresponding algorithms to explore its performance fully.
Although the solution speed of neural network methods is generally slower than that of traditional methods such as the finite element method, PINN has advantages in scenarios that require processing data for multiple time steps. Traditional numerical methods can only handle data for one time step at a time, which may exhibit limitations as the amount of data increases. In contrast, PINN can utilize data from multiple time steps simultaneously, demonstrating advantages and in some cases \citep{ref43}, exhibiting faster speed and higher accuracy.
\printcredits

\bibliographystyle{cas-model2-names}

\bibliography{cankao}

\begin{thebibliography}{39}
\expandafter\ifx\csname natexlab\endcsname\relax\def\natexlab#1{#1}\fi
\providecommand{\url}[1]{\texttt{#1}}
\providecommand{\href}[2]{#2}
\providecommand{\path}[1]{#1}
\providecommand{\DOIprefix}{doi:}
\providecommand{\ArXivprefix}{arXiv:}
\providecommand{\URLprefix}{URL: }
\providecommand{\Pubmedprefix}{pmid:}
\providecommand{\doi}[1]{\href{http://dx.doi.org/#1}{\path{#1}}}
\providecommand{\Pubmed}[1]{\href{pmid:#1}{\path{#1}}}
\providecommand{\bibinfo}[2]{#2}
\ifx\xfnm\relax \def\xfnm[#1]{\unskip,\space#1}\fi
\bibitem[{{Amini Niaki} et~al.(2021){Amini Niaki}, Haghighat, Campbell,
  Poursartip and Vaziri}]{ref21}
\bibinfo{author}{{Amini Niaki}, S.}, \bibinfo{author}{Haghighat, E.},
  \bibinfo{author}{Campbell, T.}, \bibinfo{author}{Poursartip, A.},
  \bibinfo{author}{Vaziri, R.}, \bibinfo{year}{2021}.
\newblock \bibinfo{title}{Physics-informed neural network for modelling the
  thermochemical curing process of composite-tool systems during manufacture}.
\newblock \bibinfo{journal}{Computer Methods in Applied Mechanics and
  Engineering} \bibinfo{volume}{384}, \bibinfo{pages}{113959}.
\newblock \URLprefix
  \url{https://www.sciencedirect.com/science/article/pii/S0045782521002966},
  \DOIprefix\doi{https://doi.org/10.1016/j.cma.2021.113959}.
\bibitem[{Bandai and Ghezzehei(2021)}]{ref15}
\bibinfo{author}{Bandai, T.}, \bibinfo{author}{Ghezzehei, T.A.},
  \bibinfo{year}{2021}.
\newblock \bibinfo{title}{Physics-informed neural networks with monotonicity
  constraints for richardson-richards equation: Estimation of constitutive
  relationships and soil water flux density from volumetric water content
  measurements}.
\newblock \bibinfo{journal}{Water Resources Research} \bibinfo{volume}{57}.
\bibitem[{Barreau et~al.(2021)Barreau, Aguiar, Liu and Johansson}]{ref42}
\bibinfo{author}{Barreau, M.}, \bibinfo{author}{Aguiar, M.},
  \bibinfo{author}{Liu, J.}, \bibinfo{author}{Johansson, K.H.},
  \bibinfo{year}{2021}.
\newblock \bibinfo{title}{Physics-informed learning for identification and
  state reconstruction of traffic density}, in: \bibinfo{booktitle}{2021 60th
  IEEE Conference on Decision and Control (CDC)}, pp.
  \bibinfo{pages}{2653--2658}.
\newblock \DOIprefix\doi{10.1109/CDC45484.2021.9683295}.
\bibitem[{Bi et~al.(2023)Bi, Xie, Zhang, Chen, Gu and Tian}]{ref43}
\bibinfo{author}{Bi, K.}, \bibinfo{author}{Xie, L.}, \bibinfo{author}{Zhang,
  H.}, \bibinfo{author}{Chen, X.}, \bibinfo{author}{Gu, X.},
  \bibinfo{author}{Tian, Q.}, \bibinfo{year}{2023}.
\newblock \bibinfo{title}{Accurate medium-range global weather forecasting with
  3d neural networks}.
\newblock \bibinfo{journal}{Nature} , \bibinfo{pages}{1--6}.
\bibitem[{Chen et~al.(2021)Chen, Huang, Zhang, Zeng, Wang, Zhang and
  Yan}]{ref33}
\bibinfo{author}{Chen, Y.}, \bibinfo{author}{Huang, D.},
  \bibinfo{author}{Zhang, D.}, \bibinfo{author}{Zeng, J.},
  \bibinfo{author}{Wang, N.}, \bibinfo{author}{Zhang, H.},
  \bibinfo{author}{Yan, J.}, \bibinfo{year}{2021}.
\newblock \bibinfo{title}{Theory-guided hard constraint projection (hcp): A
  knowledge-based data-driven scientific machine learning method}.
\newblock \bibinfo{journal}{Journal of Computational Physics}
  \bibinfo{volume}{445}, \bibinfo{pages}{110624}.
\newblock \URLprefix
  \url{https://www.sciencedirect.com/science/article/pii/S0021999121005192},
  \DOIprefix\doi{https://doi.org/10.1016/j.jcp.2021.110624}.
\bibitem[{Codina et~al.(1998)Codina, V{\'a}zquez and Zienkiewicz}]{ref10}
\bibinfo{author}{Codina, R.}, \bibinfo{author}{V{\'a}zquez, M.},
  \bibinfo{author}{Zienkiewicz, O.C.}, \bibinfo{year}{1998}.
\newblock \bibinfo{title}{A general algorithm for compressible and
  incompressible flows. part iii: The semi-implicit form}.
\newblock \bibinfo{journal}{International Journal for Numerical Methods in
  Fluids} \bibinfo{volume}{27}, \bibinfo{pages}{13--32}.
\bibitem[{Dulebenets(2021)}]{ref36}
\bibinfo{author}{Dulebenets, M.A.}, \bibinfo{year}{2021}.
\newblock \bibinfo{title}{An adaptive polyploid memetic algorithm for
  scheduling trucks at a cross-docking terminal}.
\newblock \bibinfo{journal}{Information Sciences} \bibinfo{volume}{565},
  \bibinfo{pages}{390--421}.
\newblock \URLprefix
  \url{https://www.sciencedirect.com/science/article/pii/S002002552100181X},
  \DOIprefix\doi{https://doi.org/10.1016/j.ins.2021.02.039}.
\bibitem[{Dwivedi and Srinivasan(2020)}]{ref20}
\bibinfo{author}{Dwivedi, V.}, \bibinfo{author}{Srinivasan, B.},
  \bibinfo{year}{2020}.
\newblock \bibinfo{title}{Physics informed extreme learning machine (pielm)–a
  rapid method for the numerical solution of partial differential equations}.
\newblock \bibinfo{journal}{Neurocomputing} \bibinfo{volume}{391},
  \bibinfo{pages}{96--118}.
\newblock \URLprefix
  \url{https://www.sciencedirect.com/science/article/pii/S0925231219318144},
  \DOIprefix\doi{https://doi.org/10.1016/j.neucom.2019.12.099}.
\bibitem[{Ethier and Steinman(1994)}]{ref31}
\bibinfo{author}{Ethier, C.R.}, \bibinfo{author}{Steinman, D.},
  \bibinfo{year}{1994}.
\newblock \bibinfo{title}{Exact fully 3d navier--stokes solutions for
  benchmarking}.
\newblock \bibinfo{journal}{International Journal for Numerical Methods in
  Fluids} \bibinfo{volume}{19}, \bibinfo{pages}{369--375}.
\bibitem[{F.R.S.(1923)}]{ref17}
\bibinfo{author}{F.R.S., G.T.}, \bibinfo{year}{1923}.
\newblock \bibinfo{title}{Lxxv. on the decay of vortices in a viscous fluid}.
\newblock \bibinfo{journal}{The London, Edinburgh, and Dublin Philosophical
  Magazine and Journal of Science} \bibinfo{volume}{46},
  \bibinfo{pages}{671--674}.
\newblock \URLprefix \url{https://doi.org/10.1080/14786442308634295},
  \DOIprefix\doi{10.1080/14786442308634295},
  \href{http://arxiv.org/abs/https://doi.org/10.1080/14786442308634295}{\tt
  arXiv:https://doi.org/10.1080/14786442308634295}.
\bibitem[{Jagtap et~al.(2020)Jagtap, Kawaguchi and Karniadakis}]{ref9}
\bibinfo{author}{Jagtap, A.D.}, \bibinfo{author}{Kawaguchi, K.},
  \bibinfo{author}{Karniadakis, G.E.}, \bibinfo{year}{2020}.
\newblock \bibinfo{title}{Adaptive activation functions accelerate convergence
  in deep and physics-informed neural networks}.
\newblock \bibinfo{journal}{Journal of Computational Physics}
  \bibinfo{volume}{404}, \bibinfo{pages}{109136}.
\bibitem[{Kani and Elsheikh(2018)}]{ref8}
\bibinfo{author}{Kani, J.N.}, \bibinfo{author}{Elsheikh, A.H.},
  \bibinfo{year}{2018}.
\newblock \bibinfo{title}{Reduced order modeling of subsurface multiphase flow
  models using deep residual recurrent neural networks}.
\newblock \bibinfo{journal}{Transport in Porous Media} .
\bibitem[{Kavoosi et~al.(2019a)Kavoosi, Dulebenets, Abioye, Pasha, Theophilus,
  Wang, Kampmann and Mikijeljevi{\'c}}]{ref37}
\bibinfo{author}{Kavoosi, M.}, \bibinfo{author}{Dulebenets, M.A.},
  \bibinfo{author}{Abioye, O.}, \bibinfo{author}{Pasha, J.},
  \bibinfo{author}{Theophilus, O.}, \bibinfo{author}{Wang, H.},
  \bibinfo{author}{Kampmann, R.}, \bibinfo{author}{Mikijeljevi{\'c}, M.},
  \bibinfo{year}{2019}a.
\newblock \bibinfo{title}{Berth scheduling at marine container terminals: A
  universal island-based metaheuristic approach}.
\newblock \bibinfo{journal}{Maritime Business Review} \bibinfo{volume}{5},
  \bibinfo{pages}{30--66}.
\bibitem[{Kavoosi et~al.(2019b)Kavoosi, Dulebenets, Abioye, Pasha, Wang and
  Chi}]{ref39}
\bibinfo{author}{Kavoosi, M.}, \bibinfo{author}{Dulebenets, M.A.},
  \bibinfo{author}{Abioye, O.F.}, \bibinfo{author}{Pasha, J.},
  \bibinfo{author}{Wang, H.}, \bibinfo{author}{Chi, H.}, \bibinfo{year}{2019}b.
\newblock \bibinfo{title}{An augmented self-adaptive parameter control in
  evolutionary computation: A case study for the berth scheduling problem}.
\newblock \bibinfo{journal}{Advanced Engineering Informatics}
  \bibinfo{volume}{42}, \bibinfo{pages}{100972}.
\newblock \URLprefix
  \url{https://www.sciencedirect.com/science/article/pii/S1474034619305452},
  \DOIprefix\doi{https://doi.org/10.1016/j.aei.2019.100972}.
\bibitem[{Liu and Wang(2019)}]{ref27}
\bibinfo{author}{Liu, D.}, \bibinfo{author}{Wang, Y.}, \bibinfo{year}{2019}.
\newblock \bibinfo{title}{Multi-fidelity physics-constrained neural network and
  its application in materials modeling}.
\newblock \bibinfo{journal}{Journal of mechanical design} ,
  \bibinfo{pages}{141}.
\bibitem[{Liu et~al.(2020)Liu, Cai and Xu}]{ref14}
\bibinfo{author}{Liu, Z.}, \bibinfo{author}{Cai, W.}, \bibinfo{author}{Xu,
  Z.Q.J.}, \bibinfo{year}{2020}.
\newblock \bibinfo{title}{Multi-scale deep neural network (mscalednn) for
  solving poisson-boltzmann equation in complex domains}.
\newblock \bibinfo{journal}{Communications in Computational Physics}
  \bibinfo{volume}{28}.
\bibitem[{L{\"o}hner et~al.(1984)L{\"o}hner, Morgan and Zienkiewicz}]{ref16}
\bibinfo{author}{L{\"o}hner, R.}, \bibinfo{author}{Morgan, K.},
  \bibinfo{author}{Zienkiewicz, O.C.}, \bibinfo{year}{1984}.
\newblock \bibinfo{title}{The solution of non-linear hyperbolic equation
  systems by the finite element method}.
\newblock \bibinfo{journal}{International Journal for Numerical Methods in
  Fluids} \bibinfo{volume}{4}, \bibinfo{pages}{1043--1063}.
\bibitem[{Long et~al.(2019)Long, Lu and Dong}]{ref6}
\bibinfo{author}{Long, Z.}, \bibinfo{author}{Lu, Y.}, \bibinfo{author}{Dong,
  B.}, \bibinfo{year}{2019}.
\newblock \bibinfo{title}{Pde-net 2.0: Learning pdes from data with a
  numeric-symbolic hybrid deep network}.
\newblock \bibinfo{journal}{Journal of Computational Physics}
  \bibinfo{volume}{399}, \bibinfo{pages}{108925}.
\bibitem[{Long et~al.(2018)Long, Lu, Ma and Dong}]{ref5}
\bibinfo{author}{Long, Z.}, \bibinfo{author}{Lu, Y.}, \bibinfo{author}{Ma, X.},
  \bibinfo{author}{Dong, B.}, \bibinfo{year}{2018}.
\newblock \bibinfo{title}{Pde-net: Learning pdes from data}, in:
  \bibinfo{booktitle}{International Conference on Machine Learning},
  \bibinfo{organization}{PMLR}. pp. \bibinfo{pages}{3208--3216}.
\bibitem[{Lyu et~al.(2022)Lyu, Zhang, Chen and Chen}]{ref32}
\bibinfo{author}{Lyu, L.}, \bibinfo{author}{Zhang, Z.}, \bibinfo{author}{Chen,
  M.}, \bibinfo{author}{Chen, J.}, \bibinfo{year}{2022}.
\newblock \bibinfo{title}{Mim: A deep mixed residual method for solving
  high-order partial differential equations}.
\newblock \bibinfo{journal}{Journal of Computational Physics}
  \bibinfo{volume}{452}, \bibinfo{pages}{110930}.
\bibitem[{Mao et~al.(2020)Mao, Jagtap and Karniadakis}]{ref29}
\bibinfo{author}{Mao, Z.}, \bibinfo{author}{Jagtap, A.D.},
  \bibinfo{author}{Karniadakis, G.E.}, \bibinfo{year}{2020}.
\newblock \bibinfo{title}{Physics-informed neural networks for high-speed
  flows}.
\newblock \bibinfo{journal}{Computer Methods in Applied Mechanics and
  Engineering} \bibinfo{volume}{360}, \bibinfo{pages}{112789}.
\newblock \URLprefix
  \url{https://www.sciencedirect.com/science/article/pii/S0045782519306814},
  \DOIprefix\doi{https://doi.org/10.1016/j.cma.2019.112789}.
\bibitem[{Pasha et~al.(2022)Pasha, Nwodu, Fathollahi-Fard, Tian, Li, Wang and
  Dulebenets}]{ref38}
\bibinfo{author}{Pasha, J.}, \bibinfo{author}{Nwodu, A.L.},
  \bibinfo{author}{Fathollahi-Fard, A.M.}, \bibinfo{author}{Tian, G.},
  \bibinfo{author}{Li, Z.}, \bibinfo{author}{Wang, H.},
  \bibinfo{author}{Dulebenets, M.A.}, \bibinfo{year}{2022}.
\newblock \bibinfo{title}{Exact and metaheuristic algorithms for the vehicle
  routing problem with a factory-in-a-box in multi-objective settings}.
\newblock \bibinfo{journal}{Advanced Engineering Informatics}
  \bibinfo{volume}{52}, \bibinfo{pages}{101623}.
\newblock \URLprefix
  \url{https://www.sciencedirect.com/science/article/pii/S1474034622000891},
  \DOIprefix\doi{https://doi.org/10.1016/j.aei.2022.101623}.
\bibitem[{Rabbani et~al.(2022)Rabbani, Oladzad-Abbasabady and
  Akbarian-Saravi}]{ref40}
\bibinfo{author}{Rabbani, M.}, \bibinfo{author}{Oladzad-Abbasabady, N.},
  \bibinfo{author}{Akbarian-Saravi, N.}, \bibinfo{year}{2022}.
\newblock \bibinfo{title}{Ambulance routing in disaster response considering
  variable patient condition: Nsga-ii and mopso algorithms}.
\newblock \bibinfo{journal}{Journal of Industrial and Management Optimization}
  \bibinfo{volume}{18}, \bibinfo{pages}{1035--1062}.
\bibitem[{Raissi et~al.(2019)Raissi, Perdikaris and Karniadakis}]{ref30}
\bibinfo{author}{Raissi, M.}, \bibinfo{author}{Perdikaris, P.},
  \bibinfo{author}{Karniadakis, G.}, \bibinfo{year}{2019}.
\newblock \bibinfo{title}{Physics-informed neural networks: A deep learning
  framework for solving forward and inverse problems involving nonlinear
  partial differential equations}.
\newblock \bibinfo{journal}{Journal of Computational Physics}
  \bibinfo{volume}{378}, \bibinfo{pages}{686--707}.
\newblock \URLprefix
  \url{https://www.sciencedirect.com/science/article/pii/S0021999118307125},
  \DOIprefix\doi{https://doi.org/10.1016/j.jcp.2018.10.045}.
\bibitem[{Raissi et~al.(2017a)Raissi, Perdikaris and Karniadakis}]{ref1}
\bibinfo{author}{Raissi, M.}, \bibinfo{author}{Perdikaris, P.},
  \bibinfo{author}{Karniadakis, G.E.}, \bibinfo{year}{2017}a.
\newblock \bibinfo{title}{Physics informed deep learning (part i): Data-driven
  solutions of nonlinear partial differential equations}.
\newblock \bibinfo{journal}{arXiv preprint arXiv:1711.10561} .
\bibitem[{Raissi et~al.(2017b)Raissi, Perdikaris and Karniadakis}]{ref2}
\bibinfo{author}{Raissi, M.}, \bibinfo{author}{Perdikaris, P.},
  \bibinfo{author}{Karniadakis, G.E.}, \bibinfo{year}{2017}b.
\newblock \bibinfo{title}{Physics informed deep learning (part ii): Data-driven
  discovery of nonlinear partial differential equations}.
\newblock \bibinfo{journal}{arXiv preprint arXiv:1711.10566} .
\bibitem[{Raissi et~al.(2020)Raissi, Yazdani and Karniadakis}]{ref4}
\bibinfo{author}{Raissi, M.}, \bibinfo{author}{Yazdani, A.},
  \bibinfo{author}{Karniadakis, G.E.}, \bibinfo{year}{2020}.
\newblock \bibinfo{title}{Hidden fluid mechanics: Learning velocity and
  pressure fields from flow visualizations}.
\newblock \bibinfo{journal}{Science} \bibinfo{volume}{367},
  \bibinfo{pages}{1026--1030}.
\bibitem[{Ranade et~al.(2021)Ranade, Hill and Pathak}]{ref19}
\bibinfo{author}{Ranade, R.}, \bibinfo{author}{Hill, C.},
  \bibinfo{author}{Pathak, J.}, \bibinfo{year}{2021}.
\newblock \bibinfo{title}{Discretizationnet: A machine-learning based solver
  for navier–stokes equations using finite volume discretization}.
\newblock \bibinfo{journal}{Computer Methods in Applied Mechanics and
  Engineering} \bibinfo{volume}{378}, \bibinfo{pages}{113722}.
\newblock \URLprefix
  \url{https://www.sciencedirect.com/science/article/pii/S004578252100058X},
  \DOIprefix\doi{https://doi.org/10.1016/j.cma.2021.113722}.
\bibitem[{Schiassi et~al.(2022)Schiassi, D’Ambrosio, Drozd, Curti and
  Furfaro}]{ref41}
\bibinfo{author}{Schiassi, E.}, \bibinfo{author}{D’Ambrosio, A.},
  \bibinfo{author}{Drozd, K.}, \bibinfo{author}{Curti, F.},
  \bibinfo{author}{Furfaro, R.}, \bibinfo{year}{2022}.
\newblock \bibinfo{title}{Physics-informed neural networks for optimal planar
  orbit transfers}.
\newblock \bibinfo{journal}{Journal of Spacecraft and Rockets}
  \bibinfo{volume}{59}, \bibinfo{pages}{834--849}.
\bibitem[{Shankar and Deshpande(2000)}]{ref18}
\bibinfo{author}{Shankar, P.N.}, \bibinfo{author}{Deshpande, M.},
  \bibinfo{year}{2000}.
\newblock \bibinfo{title}{Fluid mechanics in the driven cavity}.
\newblock \bibinfo{journal}{Annual Review of Fluid Mechanics}
  \bibinfo{volume}{32}.
\bibitem[{Snaiki(2019)}]{ref28}
\bibinfo{author}{Snaiki, R.}, \bibinfo{year}{2019}.
\newblock \bibinfo{title}{Knowledge-enhanced deep learning for simulation of
  tropical cyclone boundary-layer winds}.
\newblock \bibinfo{journal}{Journal of Wind Engineering and Industrial
  Aerodynamics: The Journal of the International Association for Wind
  Engineering} \bibinfo{volume}{194}.
\bibitem[{Wang et~al.(2020)Wang, Zhang and Cai}]{ref13}
\bibinfo{author}{Wang, B.}, \bibinfo{author}{Zhang, W.}, \bibinfo{author}{Cai,
  W.}, \bibinfo{year}{2020}.
\newblock \bibinfo{title}{Multi-scale deep neural network (mscalednn) methods
  for oscillatory stokes flows in complex domains}.
\newblock \bibinfo{journal}{Communications in Computational Physics}
  \bibinfo{volume}{28}, \bibinfo{pages}{2139--2157}.
\bibitem[{Wang et~al.(2022)Wang, Liu and Wang}]{ref7}
\bibinfo{author}{Wang, H.}, \bibinfo{author}{Liu, Y.}, \bibinfo{author}{Wang,
  S.}, \bibinfo{year}{2022}.
\newblock \bibinfo{title}{Dense velocity reconstruction from particle image
  velocimetry/particle tracking velocimetry using a physics-informed neural
  network}.
\newblock \bibinfo{journal}{Physics of fluids} , \bibinfo{pages}{34}.
\bibitem[{Wang et~al.(2021)Wang, Wang and Perdikaris}]{ref23}
\bibinfo{author}{Wang, S.}, \bibinfo{author}{Wang, H.},
  \bibinfo{author}{Perdikaris, P.}, \bibinfo{year}{2021}.
\newblock \bibinfo{title}{On the eigenvector bias of fourier feature networks:
  From regression to solving multi-scale pdes with physics-informed neural
  networks}.
\newblock \bibinfo{journal}{Computer Methods in Applied Mechanics and
  Engineering} \bibinfo{volume}{384}, \bibinfo{pages}{113938}.
\newblock \URLprefix
  \url{https://www.sciencedirect.com/science/article/pii/S0045782521002759},
  \DOIprefix\doi{https://doi.org/10.1016/j.cma.2021.113938}.
\bibitem[{Xu et~al.(2020)Xu, Zhang, Luo, Xiao and Ma}]{ref12}
\bibinfo{author}{Xu, Z.}, \bibinfo{author}{Zhang, Y.}, \bibinfo{author}{Luo,
  T.}, \bibinfo{author}{Xiao, Y.}, \bibinfo{author}{Ma, Z.},
  \bibinfo{year}{2020}.
\newblock \bibinfo{title}{Frequency principle: Fourier analysis sheds light on
  deep neural networks}.
\newblock \bibinfo{journal}{Communications in Computational Physics} .
\bibitem[{Xu et~al.(2019)Xu, Zhang and Xiao}]{ref11}
\bibinfo{author}{Xu, Z.Q.J.}, \bibinfo{author}{Zhang, Y.},
  \bibinfo{author}{Xiao, Y.}, \bibinfo{year}{2019}.
\newblock \bibinfo{title}{Training behavior of deep neural network in frequency
  domain}, in: \bibinfo{editor}{Gedeon, T.}, \bibinfo{editor}{Wong, K.W.},
  \bibinfo{editor}{Lee, M.} (Eds.), \bibinfo{booktitle}{Neural Information
  Processing}, \bibinfo{publisher}{Springer International Publishing},
  \bibinfo{address}{Cham}. pp. \bibinfo{pages}{264--274}.
\bibitem[{Yang et~al.(2021)Yang, Meng and Karniadakis}]{ref26}
\bibinfo{author}{Yang, L.}, \bibinfo{author}{Meng, X.},
  \bibinfo{author}{Karniadakis, G.E.}, \bibinfo{year}{2021}.
\newblock \bibinfo{title}{B-pinns: Bayesian physics-informed neural networks
  for forward and inverse pde problems with noisy data}.
\newblock \bibinfo{journal}{Journal of Computational Physics}
  \bibinfo{volume}{425}, \bibinfo{pages}{109913}.
\newblock \URLprefix
  \url{https://www.sciencedirect.com/science/article/pii/S0021999120306872},
  \DOIprefix\doi{https://doi.org/10.1016/j.jcp.2020.109913}.
\bibitem[{Zhao and Zhang(2020)}]{ref35}
\bibinfo{author}{Zhao, H.}, \bibinfo{author}{Zhang, C.}, \bibinfo{year}{2020}.
\newblock \bibinfo{title}{An online-learning-based evolutionary many-objective
  algorithm}.
\newblock \bibinfo{journal}{Information Sciences} \bibinfo{volume}{509},
  \bibinfo{pages}{1--21}.
\newblock \URLprefix
  \url{https://www.sciencedirect.com/science/article/pii/S0020025519308187},
  \DOIprefix\doi{https://doi.org/10.1016/j.ins.2019.08.069}.
\bibitem[{Zienkiewicz et~al.(2014)Zienkiewicz, Taylor and Nithiarasu}]{ref25}
\bibinfo{editor}{Zienkiewicz, O.}, \bibinfo{editor}{Taylor, R.},
  \bibinfo{editor}{Nithiarasu, P.} (Eds.), \bibinfo{year}{2014}.
\newblock \bibinfo{title}{The Characteristic-Based Split (CBS) Algorithm}.
  \bibinfo{edition}{seventh edition} ed..
  \bibinfo{publisher}{Butterworth-Heinemann}, \bibinfo{address}{Oxford}.
\newblock pp. \bibinfo{pages}{87--118}.
\newblock \URLprefix
  \url{https://www.sciencedirect.com/science/article/pii/B9781856176354000340},
  \DOIprefix\doi{https://doi.org/10.1016/C2009-0-26328-8}.

\end{thebibliography}

\end{document}